\documentclass[aps,prb,notitlepage]{revtex4-1}

\usepackage{amssymb,amsmath}
\usepackage{graphicx}
\usepackage[colorlinks=true, allcolors=blue]{hyperref}

\def\[{\begin{equation}}
\def\]{\end{equation}}
\def\_#1{{\bf #1}}
\def\=#1{\overline{\overline{#1}}}
\def\o{\omega}
\def\.{\cdot}
\def\x{\times}
\def\E{\varepsilon}
\def\M{\mu}
\newcommand{\ds}{\displaystyle}
\def\l#1{\label{eq:#1}}
\def\r#1{(\ref{eq:#1})}

\def\d{\partial}

\def\ds{\displaystyle}

\renewcommand\Re{{\rm Re}}
\renewcommand\Im{{\rm Im}}

\newcommand\matr[1]{\left(\begin{array}{ccc}#1\end{array}\right)}

\begin{document}

\title{Superabsorbing metamaterial wormhole:\\
  Physical modeling and wave interaction effects}
\date{\today}
\author{Stanislav I.\ Maslovski} \email{stas@co.it.pt}
\author{Hugo R.\ L.\ Ferreira}
\author{Iurii O.\ Medvedev}
\affiliation{Instituto de
  Telecomunica\c{c}\~{o}es (Coimbra)\\
  DEEC FCTUC P\'olo II - Pinhal de Marrocos\\
  3030--290 Coimbra, Portugal}
\author{Nuno G.\ B.\ Br\'{a}s}
\affiliation{Instituto de Telecomunica\c{c}\~{o}es (Lisboa)\\
  Instituto Superior T\'{e}cnico, Av.\ Rovisco Pais 1\\
  1049--001 Lisboa, Portugal,}
\affiliation{Universidade Aut\'{o}noma de Lisboa\\
  Rua Santa Marta 56 -- Pal\'{a}cio Dos Condes Do Redondo\\
  1169-023 Lisboa, Portugal}

\begin{abstract}  
  Conjugate-impedance matched superabsorbers are metamaterial bodies
  whose effective absorption cross section greatly exceeds their
  physical dimension. Such objects are able to receive radiation when
  it is not directly incident on their surface. Here, we develop methods of
  physical modeling of such structures and investigate interactions of
  the superabsorbers with passing electromagnetic radiation. The
  particular superabsorbing structure under study is a wormhole
  comprised of meshes of loaded transmission lines. A theory of
  electromagnetic wave propagation and absorption in such metamaterial
  structures is developed. At the frequency of operation, the
  structure exhibits greatly enhanced absorption as compared to the black
  body-type absorber of the same size.  Peculiar wave absorption
  effects such as trapping of nearby passing beams of electromagnetic
  radiation are demonstrated by numerical simulations. Possible
  modifications of the wormhole structure under the goal of optimizing
  absorption while minimizing complexity of the involved
  metamaterials are discussed. Conjugate-impedance matched
  superabsorbers may find applications as efficient harvesters of
  electromagnetic radiation, novel antennas, and sensors.
\end{abstract}

\maketitle

\section{Introduction}

From wave optics, it is known that the scattering and absorption cross
sections of resonant particles can be much greater than that of
non-resonant bodies with the same dimensions.~\cite{Bohren} For
instance, the extinction cross-section in subwavelength particles
exhibiting plasmonic or polaritonic resonances can be orders of
magnitude greater than the same for a black-body type absorber of a
comparable physical
size.\cite{Tretyakov,Tribelsky,Fan,Tribelsky2,Ruan1,Ruan2,Estakhri,Miroshnichenko}
Effectively, such resonant particles are able to collect the incident
wave power from an area much bigger than their physical cross section.

The same physical principle of optimal resonant absorption is used
when designing compact receiving antennas. From the theory of wire
antennas\cite{Schelkunoff,dipole2003,dipole2010} it is known that a
short wire dipole (with length much smaller than half wavelength) is a
rather ineffective receiver unless it is loaded with complex impedance
$Z_{\rm load}(\o) = Z_{\rm dip}^*(\o)$, where $Z_{\rm dip}^*(\o)$ is
the complex conjugate of the input impedance of the dipole antenna at
the frequency $\o$. Such a conjugate-impedance matched load
compensates for the excess reactance of the short dipole antenna,
tunes it in resonance with the incident field, and provides for the
maximum of the received power.\cite{Schelkunoff}

The ultimate limit for the effective receiving area of a resonant
dipole is $(3/8\pi)\lambda^2$ (e.g., Ref.~\onlinecite{Tretyakov}),
where $\lambda$ is the radiation wavelength. Note that this limit is
determined by the wavelength rather than by the dimensions of the
dipole. If a particle supports higher order multipolar resonances
(besides the main electric dipolar mode) at the same frequency $\o$,
its absorption cross section can be made larger than that of a
resonant dipole.~\cite{Estakhri,Miroshnichenko} In fact, it can be
shown that there is no ultimate upper limit on the effective
absorption cross section of a resonant object when more and more
multipolar modes of the object pile up at the same resonant
wavelength. Analogous results for the gain and directivity of
conjugate-impedance matched antennas have been known for a long
time.~\cite{Chu,Harrington}

Cylindrical metamaterial superabsorbers utilizing isotropic
double-negative metamaterials are known from literature.~\cite{Ng}
However, their performance is limited to the normal incidence of
vertically polarized waves. Perfectly conjugate-impedance matched
metamaterials that enable optimal absorption of the incident
electromagnetic radiation in arbitrary excitation scenarios have been
proposed in Ref.~\onlinecite{eesink}. With this principle, a spherical
object --- ``metamaterial thermal black hole'' --- formed by a medium
with simultaneously negative permittivity and
permeability\cite{Veselago} can be constructed to posses arbitrarily
large absorption cross section, theoretically, independently of the
physical radius of the object.\cite{mmsuper}
The required condition is that the double-negative (DNG)
metamaterials with arbitrarily small loss and arbitrarily large ranges
of permittivity and permeability values are attainable.  Such
metamaterial superabsorbers, if realized in practice, could be used,
for example, as efficient harvesters of electromagnetic radiation at
microwave frequencies or as super-Planckian radiative heat emitters at
infrared and optical frequencies.~\cite{mmsuper,SPIE}

It is also known that the subwavelength superscattering
objects~\cite{Tribelsky} under plane wave incidence and the
metamaterial thermal black holes with optically large
radius~\cite{mmsuper,meta2016} when illuminated by Gaussian beams
exhibit very peculiar behaviors of the Poynting vectors in their
vicinity, which so far has been only studied theoretically. Unusual
wave effects, such as trapping of nearby passing beams of light, have
been predicted for such objects.~\cite{meta2016}
Although one may introduce an effective optical ``Schwarzschild radius'' for such
superabsorbers,~\cite{meta2016} they do not actually mimic the
behavior of light close to a celestial black hole.
Note that the beam trapping effects discussed here are very different from the ones
reported for the optical black holes with positive index of refraction
and the metamaterial black holes,\cite{Narimanov,Cheng,Chen,Yang}
because, in the latter, a beam has to enter the gradient
index medium in order to be captured, while, in our case, the DNG
metamaterial body captures the beam propagating in {\it free space}
nearby the body. In other words, the effective absorption cross
section of the optical black holes with positive refractive index does
not exceed their physical cross section, in contrast to the
superabsorbing structures studied in the present article. It is,
therefore, of a great interest to identify ways of physical (both
numerical and experimental) modeling of such structures.

In this article, we look for ways of such physical
modeling by employing topological equivalence (i.e. homeomorphism)
between a superabsorbing conjugate impedance matched object in
$n$-dimensional space and a corresponding wormhole structure in
$n+1$ dimensions (which is discussed in Sec.~\ref{summary}), so
that, for instance, the wave absorption and scattering effects on a
circular superabsorber in two dimensions can be modeled as for surface
waves on an equivalent wormhole structure in three dimensions. The
respective wormhole structure can be realized at microwave frequencies
with readily available techniques.

This article is structured as follows.
In Sec.~\ref{summary}, a summary of the known results regarding superabsorbers such as the
metamaterial thermal black holes is given, including the details of
the coordinate transformations that establish the topological
equivalence mentioned in the previous paragraph. In
Sec.~\ref{notessec}, possible approaches for physical modeling of such
superabsorbing structures utilizing the wormhole topology are
discussed, with the most feasible one being selected as based on
meshes of loaded transmission lines, which is supported by the
analytical theory for the electromagnetic Bloch waves in the periodic
meshes of such lines developed in the same section. In
Sec.~\ref{wormholesec} the wormhole structure realization, its
theoretical analysis, the simulation methods, and the numerical results
are presented (with the implementation details of the simulation
methods given in the appendix).
Finally, in Sec.~\ref{concl} the main conclusions of the present study are drawn.

\section{\label{summary}Known spherical and cylindrical metamaterial superabsorbers}

From our previous studies,~\cite{mmsuper} we know that a spherical metamaterial body
with radius $a$ and radially dependent isotropic complex permittivity
$\E(r) = \E'(r)-j\E''(r)$ and permeability $\M(r) = \M'(r)-j\M''(r)$
[at some frequency~$\o$; the time dependence is assumed to be of
the form $\exp(+j\o t)$, where $j=\sqrt{-1}$] satisfying
\[
  {\E'(r)\over\E_0} = {\M'(r)\over\M_0} = -{a^2\over r^2},
  \l{epsmu}
\]
\[
  \left|{\E''(r)\over\E'(r)}\right| = \left|{\M''(r)\over\M'(r)}\right| = \tan\delta\rightarrow 0,
  \l{tand}
\]
has the effective absorption cross-section
$\sigma_{\rm abs}(\o)\rightarrow \infty$ at the frequency $\o$,
independently of the physical radius of the body $a$. In these
relations, $\E_0$ and $\M_0$ are the permittivity and the permeability
of the surrounding space, e.g.\ free space. The parameters $\E_0$ and
$\M_0$ are assumed real-valued.

For objects made of the materials with finite values of the loss tangent:
$\tan\delta > 0$, and a limited variation range of the relative material
parameters when $r\rightarrow 0$: $|{\E'(r)/\E_0}| < \infty$,
$|{\M'(r)/\M_0}| < \infty$, the absorption cross section is finite,
but still it can be large as compared to the physical dimensions of
the body: $\sigma_{\rm abs}\gg\pi a^2$, even when $a \gg \lambda$,
where $\lambda = 2\pi c/\o$ is the radiation wavelength (with
$c = 1/\sqrt{\E_0\M_0}$ being the speed of light in the surrounding
space).

This result can be explained by the fact that a body with the
parameters~\r{epsmu} and \r{tand} is conjugate-impedance matched with the
surrounding space, at every possible spatial harmonic of the incident
field.~\cite{mmsuper} In order to prove this fact one can use, for
instance, the expansion of an arbitrary incident plane wave into
spherical harmonics with the origin at the center of the body, as it
was done in Ref.~\onlinecite{mmsuper}. It can be shown that {\em all}
incident spherical harmonics in this expansion are absorbed by such
body without reflections, and thus the body is able to receive {\em
  all} (theoretically, infinite) power transported by a plane
wave. This result means that the absorption cross section of such
a body is also theoretically infinite.

An analogous result can be obtained as well in the case of a
cylindrical body~\cite{eesink} with radius $a$ and anisotropic
radially dependent material parameters
\[
  {\=\E'\over\E_0} = {\=\M'\over\M_0} = \matr{
    -m &  0          & 0 \\
    0  & -{1\over m} & 0 \\
    0  &  0          & -{1\over m}\left({a^2\over\rho^2}\right)^{1+1/m}\!\!\!},
  \l{epsmucyl}
\]
where the components of the material tensors are given in the
cylindrical coordinate system
$(x_1,x_2,x_3) \equiv (\rho, \varphi, z)$, with $\rho$ being the
radial distance in this system, and $m > 0$ being an arbitrary
parameter. When $m = 1$, the material parameters \r{epsmucyl} become
uniaxial with respect to the $z$-axis, and thus are isotropic in the
$xy$-plane. Because the latter case is simpler to realize in practice,
in what follows, we select $m = 1$.

The superabsorbing property of such spherical and cylindrical objects
can be also explained with a coordinate transformation (transformation
optics~\cite{Pendry}) technique.~\cite{mmsuper,eesink} Namely, under
the coordinate transformation $r\mapsto a^2/r$ (in the spherical case)
or $\rho\mapsto a^{m+1}/\rho^m$ (in the cylindrical case), the media
with the parameters~\r{epsmu} and~\r{epsmucyl} transform into a
uniform DNG medium with isotropic parameters $\E' = -\E_0$ and
$\M' = -\M_0$, i.e., they transform into the
left-handed~\cite{Veselago} counterpart of the surrounding space. The
same transformation also maps the region $r < a$ into the region
$r > a$ (or $\rho < a$ to $\rho > a$ in the cylindrical case), while
keeping the tangential components of the electric and magnetic fields
intact at the surface $r = a$ ($\rho = a$).

Therefore, the plane wave incidence onto a spherical or cylindrical object
with the parameters \r{epsmu} or \r{epsmucyl} in a $n$-dimensional
space (for the spherical case, $n=3$; for the cylindrical case,
$n = 2$) can be equivalently reformulated as a problem defined on a
hypersurface in $(n+1)$-dimensional space, in which the two separate
$r > a$ regions --- the surrounding space and the transformed material
object --- are joined on the same $n$-dimensional sphere, on which
$r = a$. Topologically, a junction of two such regions or subspaces is
a $(n+1)$-dimensional wormhole.

When $\tan\delta\rightarrow 0$, the material parameters $\E(\o)$ and
$\M(\o)$ in these two joined regions differ only by sign. Therefore,
any time-harmonic solution (with frequency $\o$) of the uniform Maxwell
equations in the first region has a time-reversal ``mirror'' solution
in the second region. From here it follows that if the tangential
components of $\_E$ and $\_H$ in both subspaces are
continuous at the interface $r = a$ when passing from one subspace to
another, any wave propagating from $r = \infty$ towards $r = a$ in the
first region will be continued by a wave propagating from $r = a$
towards $r = \infty$ in the second region, without any reflections at
the interface. Thus, this observation allows for an alternative explanation
of the superabsorption phenomenon and, as will be shown later, also
provides us with an ability to demonstrate the superabsorption effect
in practice.

\section{\label{notessec}Towards physical modeling of superabsorption
  effect in 2D}

Unfortunately, with the facilities that are currently at our hands,
a practical demonstration of the superabsorption effect in three
dimensions appears to be rather difficult. Therefore, here we aim at
physical modeling of this effect in a setup with reduced
dimensionality, namely, in just two dimensions (2D).

Indeed, the case of the cylindrical object mentioned in Sec.~\ref{summary}
reduces to an effectively 2D case when the wave vector of an
incident wave lies in the $xy$-plane. In this case, an incident wave
of the transverse electric (TE) polarization has the electric field
vector $\_E^{\rm inc}$ parallel to the $z$-axis, and the magnetic
field vector $\_H^{\rm inc}$ in the $xy$-plane. Conversely, an
incident wave of the transverse magnetic (TM) polarization has the magnetic
field vector along the $z$-axis and the electric field vector in the
$xy$-plane. The fields are independent of the $z$-coordinate.

Note that the $zz$-component of the permeability tensor $\=\M$ has no
influence on such waves of the TE polarization (respectively, the
$zz$-component of the permittivity tensor $\=\E$ has no effect on the
waves of the TM polarization). In particular, this means that the
superabsorption effect in the 2D case can be demonstrated for the
waves of the TE polarization by, for example, using a double negative
(DNG) metamaterial with isotropic negative permeability
(which is the case of Ref.~\onlinecite{Ng}), or even a
metamaterial which is non-magnetic along the $z$-axis. 

Moreover, for the waves of the TE polarization, only the
$zz$-component of $\=\E$ matters, because in such waves the electric
field is oriented along the $z$-axis. Hence, the in-plane $xy$
components of the dielectric tensor can have arbitrary values without
affecting the performance of the superabsorber for the incident waves
of this polarization.

Based on these considerations, one may think of using the well-known
uniaxial metamaterial designs~\cite{Padilla} when realizing the
superabsorber. In such designs, the negative dielectric permittivity
is realized with metallic rods or strips, and the negative magnetic
response is due to split ring resonators (SRR).  There is, however, a
disadvantage in these designs: because the magnetic response is
realized with a resonant inclusion (the SRR), the loss tangent of the
effective magnetic permeability of such metamaterials is relatively
high. Additionally, the realizable range of $|\M_{\rm eff}|$ in such
media is rather low: It is usually hard to obtain
$|\M_{\rm eff}/\M_0| \gtrsim 3$ while maintaining a reasonably small
$\tan\delta$ (the latter limitation is less critical in the 2D case,
because, in this case, one has to realize a material with
$\M'_{\rho}/\M_0= \M'_{\varphi}/\M_0 = -1$). These considerations have
led us to a realization based on considerably different ideas, which is
discussed in the next section.

\subsection{Superabsorbing TL-based wormhole
  structure}

\begin{figure}[htb]
  \centering
  \includegraphics[width=0.5\linewidth]{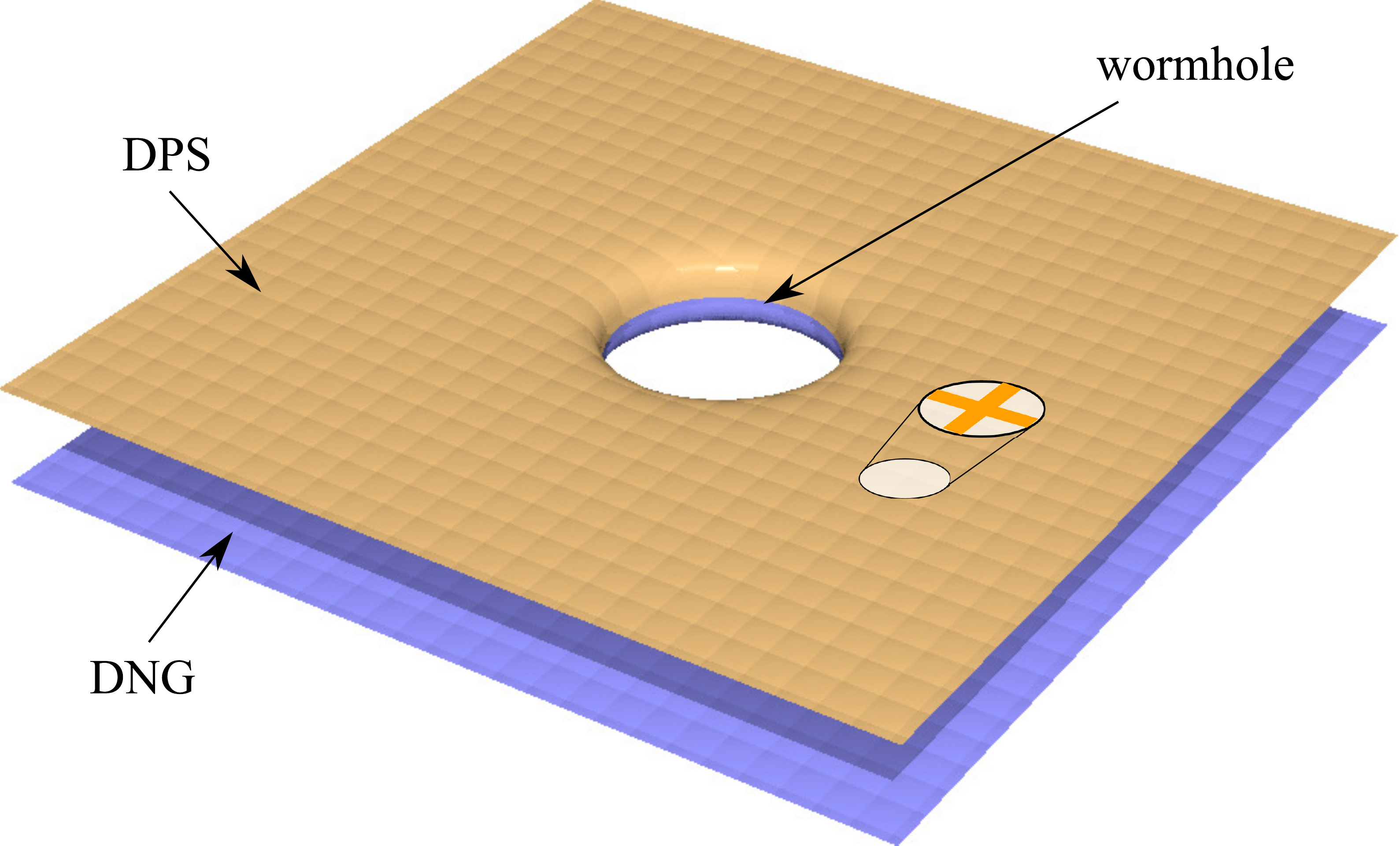}
  \caption{\label{wormhole} (Color online) Schematic representation of
    a TL-based wormhole structure formed by two electrically connected
    two-dimensional DPS and DNG domains. The DPS and DNG domains are
    realized by employing TL-based metamaterials, which are 2D meshes
    of loaded TLs. In this figure the domains are shown as golden
    (top) and blue (bottom) surfaces of zero thickness, although, in a
    real structure, the two meshes will always have finite
    thickness. The two TL meshes are electrically connected at the
    circumference $\rho = a$ of the wormhole neck.}
\end{figure}

The limitations of the conventional metamaterials discussed above can
be overcome when using 2D metamaterials realized with meshes of loaded
transmission lines (TL). The theory of such 2D metamaterials has been
developed in a number of works (for a review, see
Ref.~\onlinecite{Eleftheriades}). There is also a possibility to
extend such concepts to 3D.\cite{Grbic,Pekka,Pekka2} The
electromagnetic waves in such structures are represented by waves of
electric currents and voltages in the TL segments.

Moreover, as it was mentioned in Sec.~\ref{summary}, a
conjugate-impedance matched $n$-dimensional superabsorber can be
modeled with a wormhole structure in space with $(n+1)$
dimensions. Replacing a highly nonuniform DNG metamaterial object by
an equivalent wormhole ``tunnel'' to a uniform DNG subspace greatly
reduces the realization costs and complexity while preserving all
observable physical phenomena associated with the original structure.

In our case, the 2D superabsorption effect can be modeled by a
wormhole structure composed of two separate TL meshes: the DPS mesh
and the DNG mesh, electrically connected at the circumference of the
wormhole. This structure is shown in Fig.~\ref{wormhole}. In what
follows, we discuss the realization of such 2D metamaterials and
derive conditions under which the DPS and DNG domains in the top
and bottom halves of the structure shown in Fig.~\ref{wormhole} are
conjugate-impedance matched, which is necessary for the
superabsorption effect to occur.

\subsection{\label{TLmesh}Realizing DPS and DNG domains}

\begin{figure}[htb]
  \centering
  \includegraphics[width=0.5\linewidth]{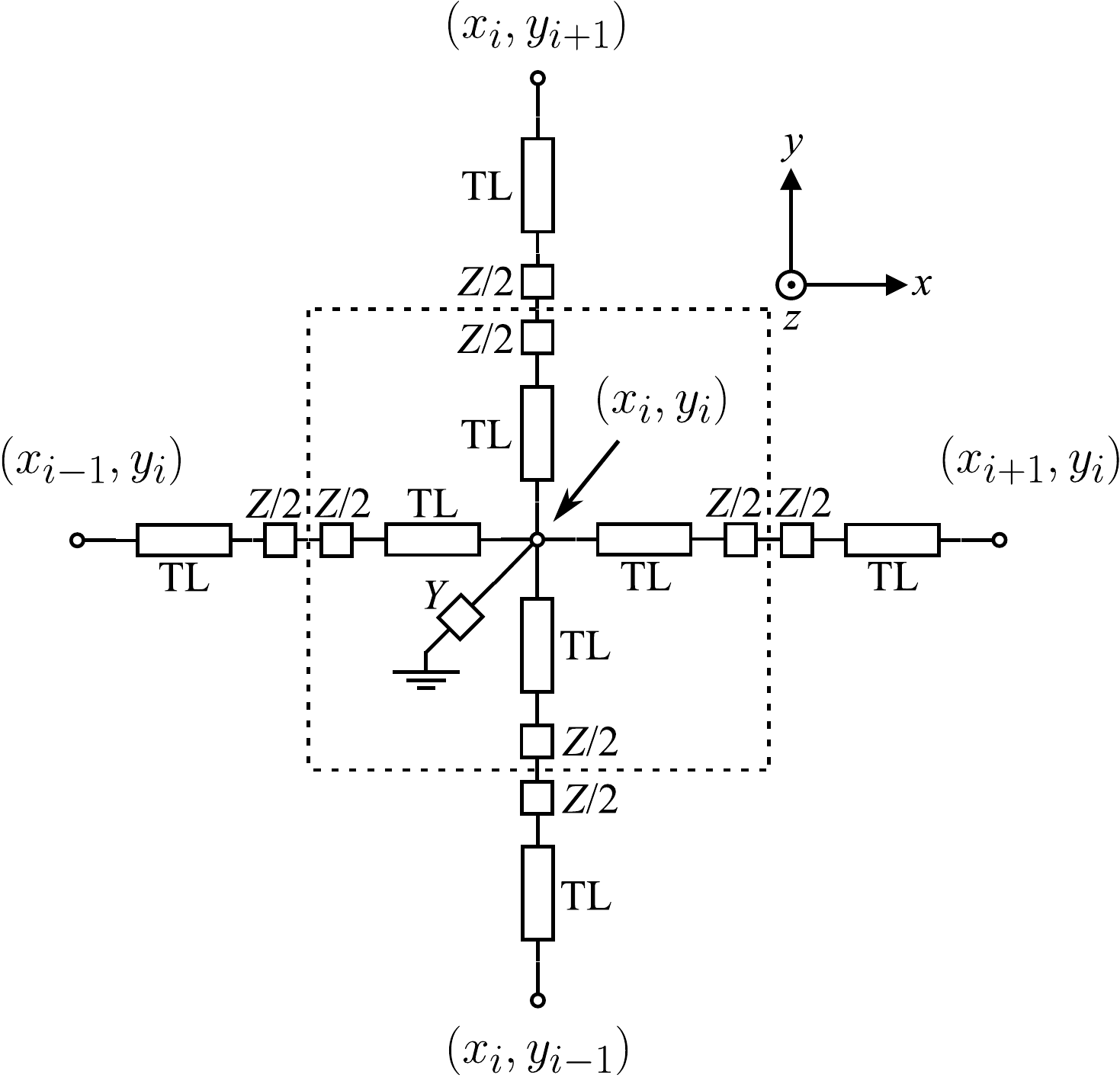}
  \caption{\label{2DTL} Circuit diagram of the loaded 2D
    transmission line (TL) mesh. The unit cell of the structure is
    indicated with a dashed square. The distance between the centers
    of the neighboring unit cells is $x_{i+1}-x_i = y_{i+1}-y_i=
    d$. Because the load impedance $Z$ at the unit cell edges is shared
    between the two neighboring cells, it is split in
    two halves in the schematic: $Z = Z/2 + Z/2$. The shunt admittance $Y$ connects
    every node of the TL mesh with the common ground shared by all TL
    segments. The elements labeled by TL are transmission line
    segments (e.g., microstrip lines) with the characteristic impedance
    $Z_0$, the propagation factor $\beta_0$, and the length $d/2$.}
\end{figure}

The unit cell of a generic 2D TL-based metamaterial is shown in
Fig.~\ref{2DTL}. By selecting proper loads, such a TL-based
metamaterial can be made to support forward or backward waves and
thus operates effectively as a double positive (DPS) medium with
$\E'_{\rm eff}> 0$ and $\M'_{\rm eff} > 0$, or as a DNG medium with
$\E'_{\rm eff}< 0$ and $\M'_{\rm eff} < 0$.

The dispersion relation for the 2D plane waves (also called Bloch
waves) in a periodic structure with the unit cell shown in
Fig.~\ref{2DTL} can be obtained with the use of the ABCD matrices of the
TL segments, in a manner analogous to what was done in
Ref.~\onlinecite{Pekka}. The result is
\begin{align}
  \l{disp}
  \cos(k_xd) + \cos(k_yd) &= {YZ\over 4}+\cos(\beta_0 d)\left(2 + {YZ\over 4}\right) + j\sin(\beta_0 d)\left({Z\over Z_0} + {YZ_0\over 2}\right).
\end{align}
Here, $d$ is the size of the (square) unit cell, $\_k = (k_x,k_y)$ is
the wave vector of the propagating Bloch wave, $\beta_0$ and $Z_0$ are
the propagation factor and the characteristic impedance in the
(unloaded) TL segments, respectively. These parameters, in
general, depend on the frequency: $\beta_0 \equiv \beta_0(\o)$,
$Z_0 \equiv Z_0(\o)$.

Considering, for instance, all Bloch waves with a fixed value of the
transverse wavenumber $k_y$, we can express the longitudinal
propagation factor $k_x$ as
\begin{align}
  \l{eqkx}
  k_x = \pm{1\over d}\cos^{-1}\left[{YZ\over 4}-\cos(k_yd) + \cos(\beta_0 d)\left(2 + {YZ\over 4}\right) + j\sin(\beta_0 d)\left({Z\over Z_0} + {YZ_0\over 2}\right)\right].
\end{align}
Here the ambiguity in sign of $k_x$ is related to the fact that in a
reciprocal structure there always exist two waves with the positive
and the negative phase velocities which satisfy the same dispersion
equation.

We can introduce an analog of the plane wave impedance for such Bloch
waves, the Bloch impedance $Z_{\rm B}$, as a ratio between the line
voltage and the $x$-directed current in the TL mesh at any fixed cell
boundary $x = \mbox{const}$ (for more details, see
Ref.~\onlinecite{Pekka}). The result is
\[
  Z_{\rm B}=\pm\left(Z_0\tan{\beta_0 d\over 2}-{jZ\over 2}\right)\cot{k_xd\over2},
  \l{Bloch}
\]
where $k_x$ is given by \r{eqkx} and the sign must be chosen so that
$\Re\,Z_{\rm B} > 0$.

Let us consider the case when the loads are such that $Z = 1/(j\o C)$
(i.e., a serial capacitor is inserted between the line ends in the
neighboring cells) and $Y = 1/(j\o L)$ (i.e., there is a shunt
inductor to the ground at every line crossing). We can also express
the TL parameters $\beta_0$ and $Z_0$ through the inductance $L_0$ and
the capacitance $C_0$ per unit length of the TL as
$\beta_0 = \o\sqrt{L_0C_0}$ and $Z_0 = \sqrt{L_0/C_0}$.

Then, in the long wavelength limit, when $|\_k|d\ll 1$ and
$\beta_0 d\ll 1$, the solution of the dispersion equation~\r{disp} can
be approximated as
\[
  k_x = \pm\sqrt{\o^2L_{\rm eff}C_{\rm eff} - k_y^2},
  \l{kxapprox}
\]
and the Bloch impedance as
\[
  Z_{\rm B}=\pm{\omega L_{\rm eff}\over k_x},
  \l{ZBapprox}
\]
where $L_{\rm eff} = L_0 - {1/(\o^2 Cd)}$ and $C_{\rm eff} = 2C_0-{1/(\o^2 Ld)}$.

The situation without loads (i.e.\ a square mesh of the unloaded TLs)
can be modeled with the same equations when $Z = 0$, $Y = 0$. In this
case, the dispersion equation~\r{disp} reduces to
\[
  \cos(k_xd)+\cos(k_yd) = 2\cos(\beta_0 d), \l{disp0}
\]
and the expression for the Bloch impedance~\r{Bloch} to
$Z_{\rm B}=\pm Z_0\tan({\beta_0 d/2})\cot({k_xd/2}). \l{Bloch0}$ From
here, in the limit when $|\_k|d\ll 1$ and $\beta_0 d\ll 1$, we obtain
\[
k_x = \pm\sqrt{2\o^2L_0C_0 - k_y^2}, \l{kxdps0}
\]
and $Z_{\rm B}=\pm{\omega L_0/k_x}.$

By comparing Eqs. \r{kxapprox} and \r{ZBapprox} with analogous
expressions for the propagation factor and the wave impedance of
the TE-polarized plane waves in continuous media, we can identify the
parameters $C_{\rm eff}$ and $L_{\rm eff}$ as the analogs of the
permittivity $\E_{\rm eff}$ and the permeability $\M_{\rm eff}$ of
such media. Respectively, in the 2D configuration we are discussing, a
TL mesh with no loading and with $C_{\rm eff} = 2C_0$, $L_{\rm eff} = L_0$,
will model the surrounding space with the parameters $\E_0$, $\M_0$
(the DPS domain), and a loaded TL mesh such that $L_{\rm eff} < 0$,
$C_{\rm eff} < 0$ will model the medium with negative material
parameters (the DNG domain).

\subsection{Conjugate-impedance matching between DPS and DNG domains}
For the superabsorption effect to appear the absorber must be
conjugate-impedance matched to the surrounding space. In order to
study the possibility to realize this condition in 2D with the
TL-based metamaterials, let us consider an interface between a DPS
halfplane and a DNG halfplane. Without any loss of generality, we may
assume that this interface is located at $x = 0$, so that the DPS
region is located at $x < 0$ and the DNG region is at $x > 0$.

In the DPS region, both group and phase velocities of a wave
propagating towards the interface (the incident wave) are positive,
and thus $\Re\, k_x^{\rm DPS} > 0$ and one has to select the positive
branch of Eq.~\r{eqkx}. In the DNG region a wave propagating away from
the interface (the transmitted wave) has positive group velocity and
negative phase velocity. Therefore, $\Re\, k_x^{\rm DNG} < 0$, and one
has to select the negative branch of Eq.~\r{eqkx}.

The Bloch impedance for the incident wave in the DPS region, therefore,
reads
\[
  Z_{\rm B}^{\rm DPS} = Z_0^{\rm DPS}\tan{\beta_0^{\rm DPS}d\over
    2}\cot{k_x^{\rm DPS}d\over 2},
\l{ZBDPS}
\]
where we select the plus branch of Eq.~\r{Bloch} in order to have
$\Re\, Z_{\rm B}^{\rm DPS} > 0$. Respectively, the Bloch impedance for
the transmitted wave in the DNG region reads
\[
  Z_{\rm B}^{\rm DNG}=\left(Z_0^{\rm DNG}\tan{\beta_0^{\rm DNG} d\over
      2} - {jZ\over 2}\right)\cot{k_x^{\rm DNG}d\over2},
\l{ZBDNG} 
\]
where the plus sign in front of Eq.~\r{Bloch} is selected because when
$L_{\rm eff} < 0$ (which holds in the DNG domain) the real part of the
parenthesized expression in Eq.~\r{ZBDNG} is negative and also
$\Re\, k_x^{\rm DNG} < 0$.

The conjugate-impedance match of the DPS and the DNG regions occurs
when
\[
  Z_{\rm B}^{\rm DNG}(\o, k_y) = \left(Z_{\rm B}^{\rm DPS}(\o,
    k_y)\right)^*.
  \l{match}
\]
This equality must hold at a given frequency $\o$ and at {\em
  arbitrary} real $k_y$ in order for the two regions to be matched for all
propagating and evanescent spatial harmonics of the incident field
with the selected frequency $\o$.

As follows from Eqs.~\r{ZBDPS} and~\r{ZBDNG}, the required matching for
arbitrary $k_y$ can be achieved only when
\[
  k_x^{\rm DNG}(\o,k_y) = -\left(k_x^{\rm DPS}(\o,k_y)\right)^*.
\l{matchkx}
\]
In this case, Eq.~\r{match} can be reduced to
\begin{align}
  \l{Zl}
  Z =
    -2j\left(\left(Z_0^{\rm DPS}\right)^*\tan{\left(\beta_0^{\rm DPS}\right)^*d\over 2}
      + Z_0^{\rm DNG}\tan{\beta_0^{\rm DNG}d\over 2}\right).
\end{align}
By substituting this expression into Eq.~\r{disp} and using
Eqs.~\r{disp0} and~\r{matchkx} we find that in order to achieve matching at all
$k_x$ the load admittance $Y$ has to be
\begin{align}
  \l{Yl}
  Y =
    -{2j\over \cos^2({\beta_0^{\rm DNG}d/2})}\left[
      \left(\sin\left(\beta_0^{\rm DPS}d\right)\over Z_0^{\rm DPS}\right)^* \!\!+
      {\sin\left(\beta_0^{\rm DNG}d\right)\over Z_0^{\rm DNG}}\right].
\end{align}

In the case when $\beta_0^{\rm DPS} = \beta_0^{\rm DNG} = \beta_0$,
$\Im\,\beta_0\rightarrow 0$, and $Z_0^{\rm DPS} = Z_0^{\rm DNG} =
Z_0$, $\Im\,Z_0 \rightarrow 0$, Eqs.~\r{Zl} and \r{Yl} reduce to
\[
  Z = -4jZ_0\tan{\beta_0d\over 2},
  \l{Zl0}
\]
\[
  Y = -{8j\over Z_0}\tan{\beta_0d\over 2}.
  \l{Yl0}
\]
When, additionally, $\beta_0d \ll 1$, we may approximate these
relations as
\[
  Z = -2jZ_0\beta_0d = -2j\o L_0d,
\]
\[
  Y = -{4j\over Z_0}\beta_0d = -4j\o C_0d.
\]
From here one can see that the conjugate-impedance matching between
a dense unloaded DPS TL mesh and a dense loaded DNG TL mesh with
negligible loss is achieved when $L_{\rm eff}^{\rm DNG} = -L_{\rm
  eff}^{\rm DPS} = -L_0$ and
$C_{\rm eff}^{\rm DNG} = -C_{\rm eff}^{\rm DPS} = -2C_0$, which is
analogous to the conjugate matching condition for
the DPS and DNG regions of continuous media.\cite{eesink}

\section{\label{wormholesec}Wormhole structure numerical modeling and analysis}

\begin{figure}[htb]
  \centering
  \includegraphics[width=0.3\linewidth]{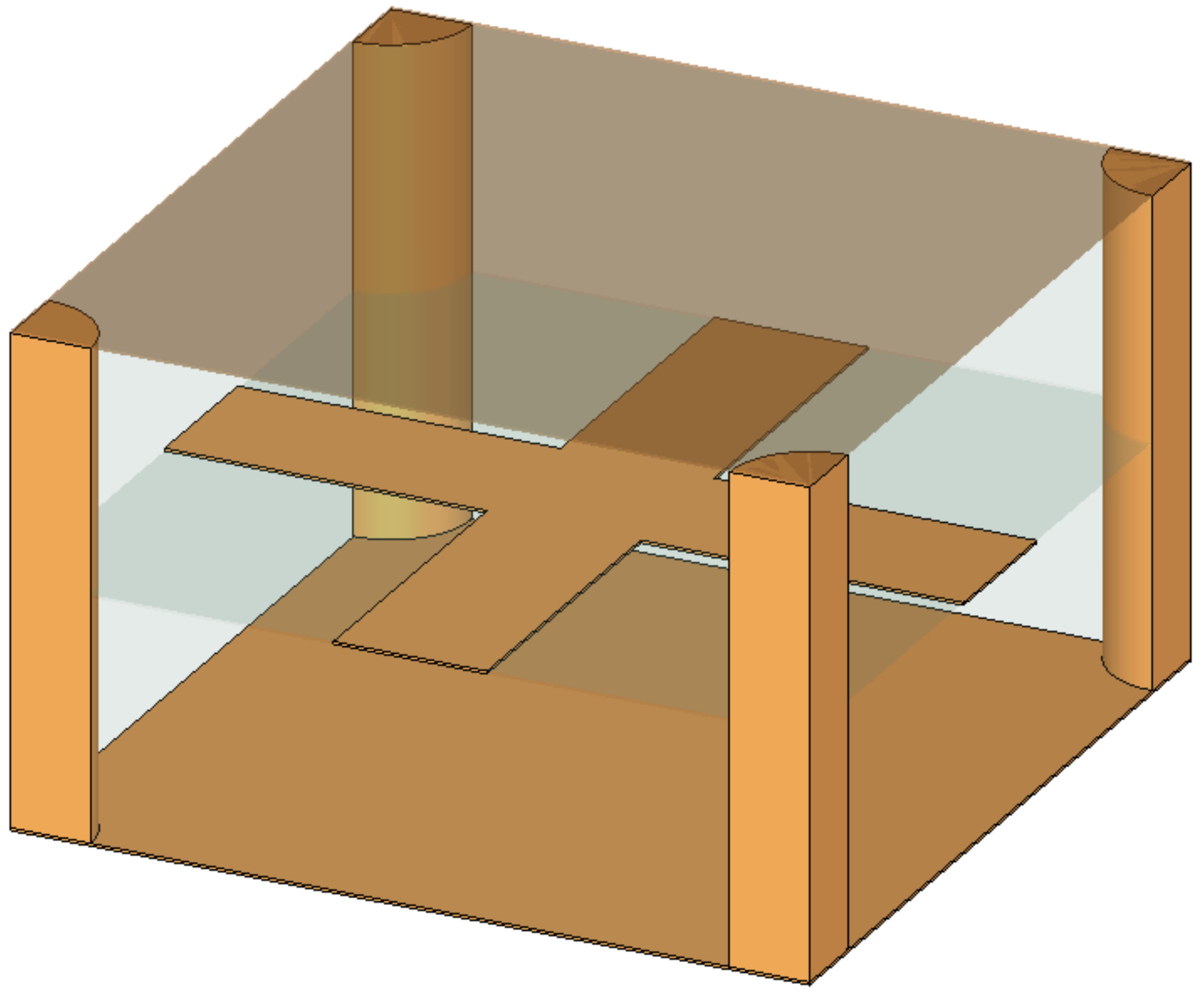}\includegraphics[width=0.3\linewidth]{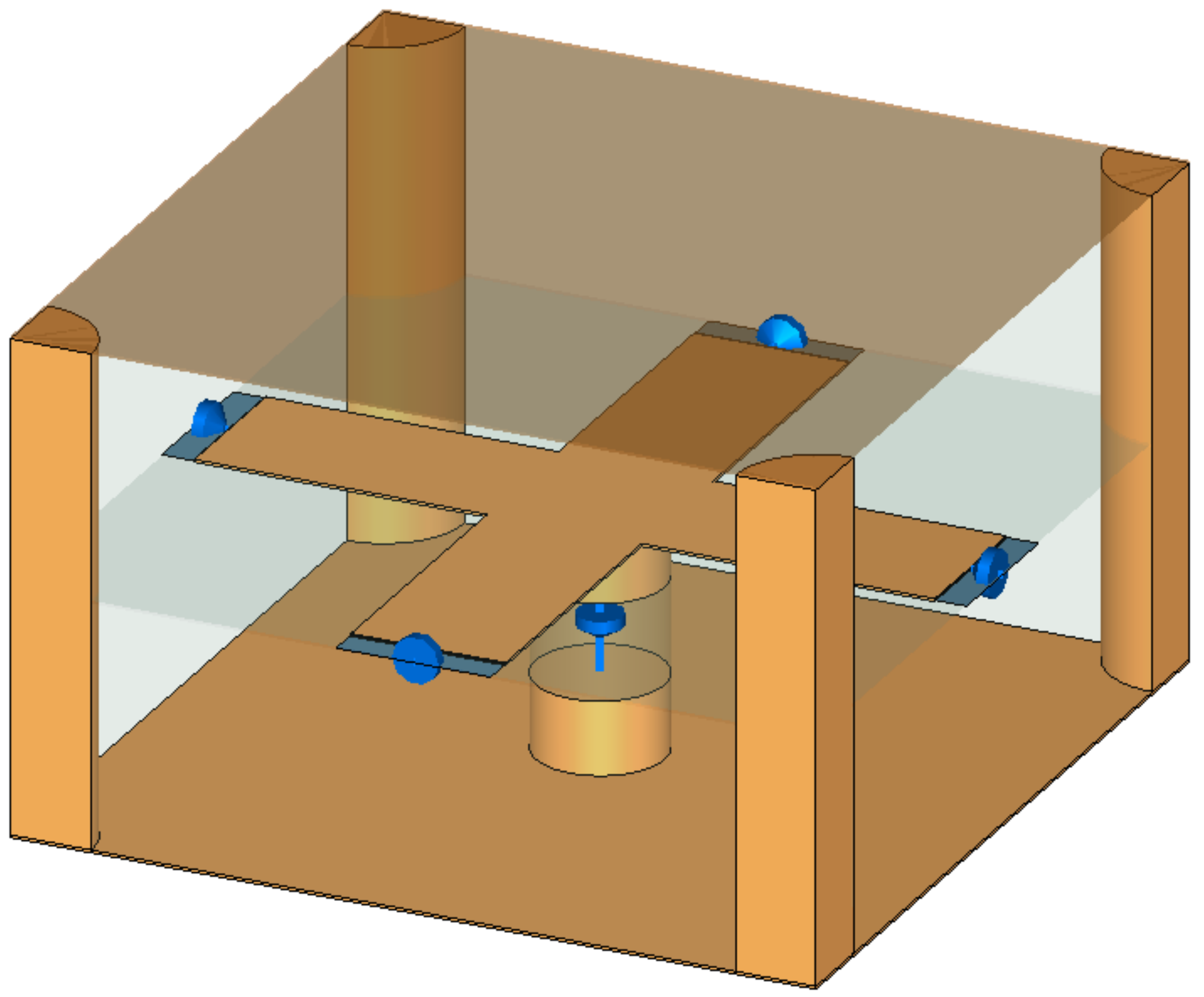}\\
  (a)\hspace{0.27\linewidth}(b)\\[1mm]
  \includegraphics[width=0.6\linewidth]{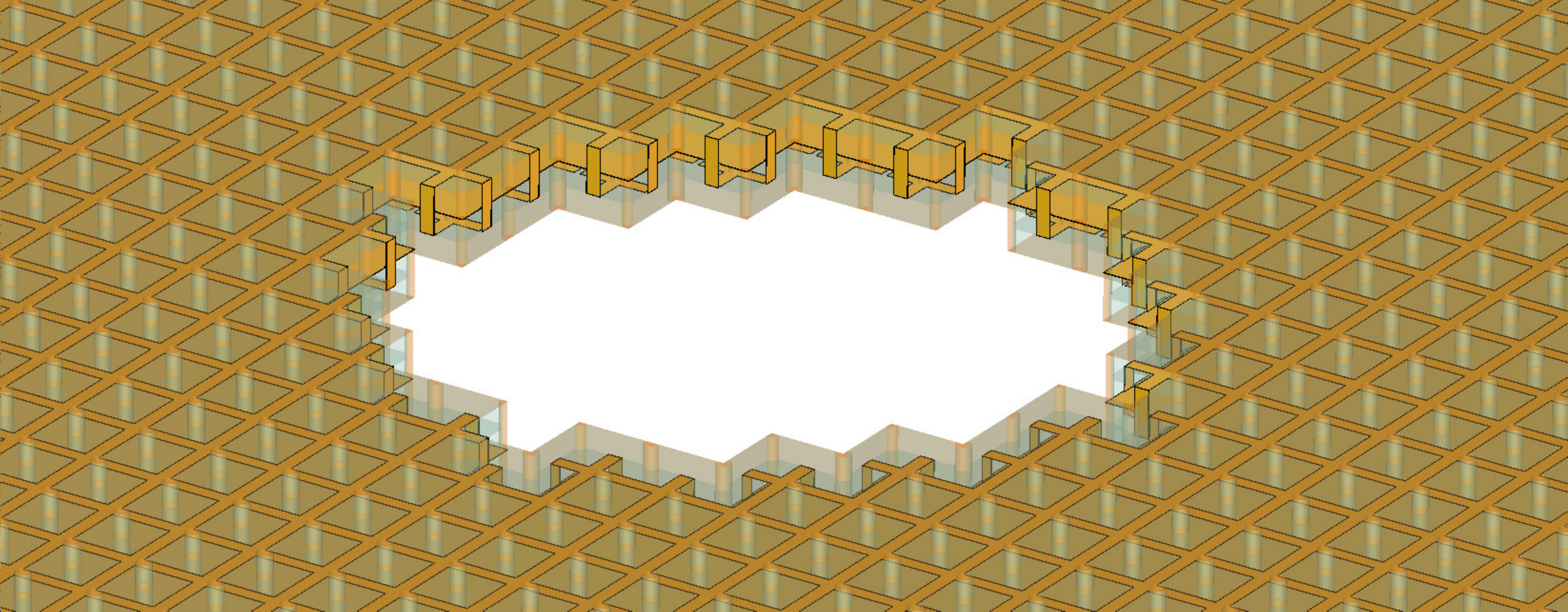}\\(c)\\
  \caption{\label{stripcell} (Color online) The unit cells of the DPS
    [panel (a)] and the DNG [panel (b)] TL meshes. The DPS unit cell
    is formed by an unloaded crossing of two symmetric strip
    lines. The DNG unit cell is formed by two crossed stripline
    segments loaded at the crossing by a lumped inductor connected to
    the ground (which realizes $Y$), and by four lumped capacitors
    at the edges of the unit cell (which realize $Z/2$). The lumped elements
    are shown in blue. The ground metalizations at the top (shown semi-transparent)
    and the bottom are electrically connected by cylindrical metallic
    pillars at the corners of the unit cell in order to prevent
    excitation of the unwanted parallel plate waveguide mode.  Panel
    (c) shows a possible realization of the wormhole structure shown in
    Fig.~\ref{wormhole} with these DPS and DNG cells.
    The top layer is formed by the DPS cells. The bottom
    layer is formed by the DNG cells. 
    The strip lines in the DPS and DNG cells are electrically
    connected at the wormhole edge.
    The pictures are produced with the CST Microwave Studio software.}
\end{figure}

In a practical realization of the wormhole structure schematically
shown in Fig.~\ref{wormhole}, one can employ meshes of strip lines
with the unit cells depicted in Fig.~\ref{stripcell}~(a,b). These unit cells
fill the DPS and the DNG domains shown in Fig.~\ref{wormhole} by the
golden (top) and the blue (bottom) surfaces, respectively. In
practice, these DPS and DNG networks can be laid atop one another so
that they are separated by the common ground plane (for example, the
bottom metalization of the DPS network can also serve as the top
metalization for the DNG network, or {\it vice versa}). Note that in
the middle of this structure, where the wormhole is located, there are
no strip lines or loads so that no in-plane
propagation may happen in that region. Instead, at the perimeter of the wormhole
neck, the open ports of the DPS unit cells are electrically connected
to the corresponding ports of the DNG unit cells with short vertical
metallic strips that pass through the opening in the middle ground
layer. Therefore, a wave propagating in the DPS network towards the
wormhole neck, after passing through the connections at the
circumference of the wormhole neck, will continue to propagate in the
DNG region in an outward direction.

In order to numerically analyze such structure composed of
many DPS and DNG unit cells, one can either use a general purpose
electromagnetic simulator (such as CST Microwave Studio\cite{CST} or
ANSYS HFSS\cite{HFSS}) or develop a custom software. However, note
that the detailed full-wave simulation of a structure with
$10^4$--$10^5$ unit cells requires a lot of computing resources.

In this work we employ a strategy in which a general purpose simulator
is only used to model isolated DPS and DNG unit cells. From these
simulations, the scattering parameters (the $S$-parameters) of the
unit cells are found. A very good approximation for the same
parameters for the main propagating mode can be also obtained
analytically by using the TL-based unit cell model from
Sec.~\ref{TLmesh}, which is done in Appendix~\ref{AppA}.

When the $S$-parameters of the cells are known, the behavior of the
whole structure formed by many thousands of cells is modeled with an
in-house simulator based on the frequency-domain transmission line
matrix (FDTLM) method.\cite{Jin} In this method, the unit cells are
represented as multiport waveguide joints or blocks with a given
number of ports (which is four for a structure with square unit cells)
and a given number of incident and reflected waves in each port (which
is two when only the main modes of a single polarization are
considered). The implementation details of the FDTLM method are
given in Appendix~\ref{AppB}.

\subsection{Numerical results of analytical model and CST simulations}

\begin{figure}[h]
  \centering
  \includegraphics[width=0.5\linewidth]{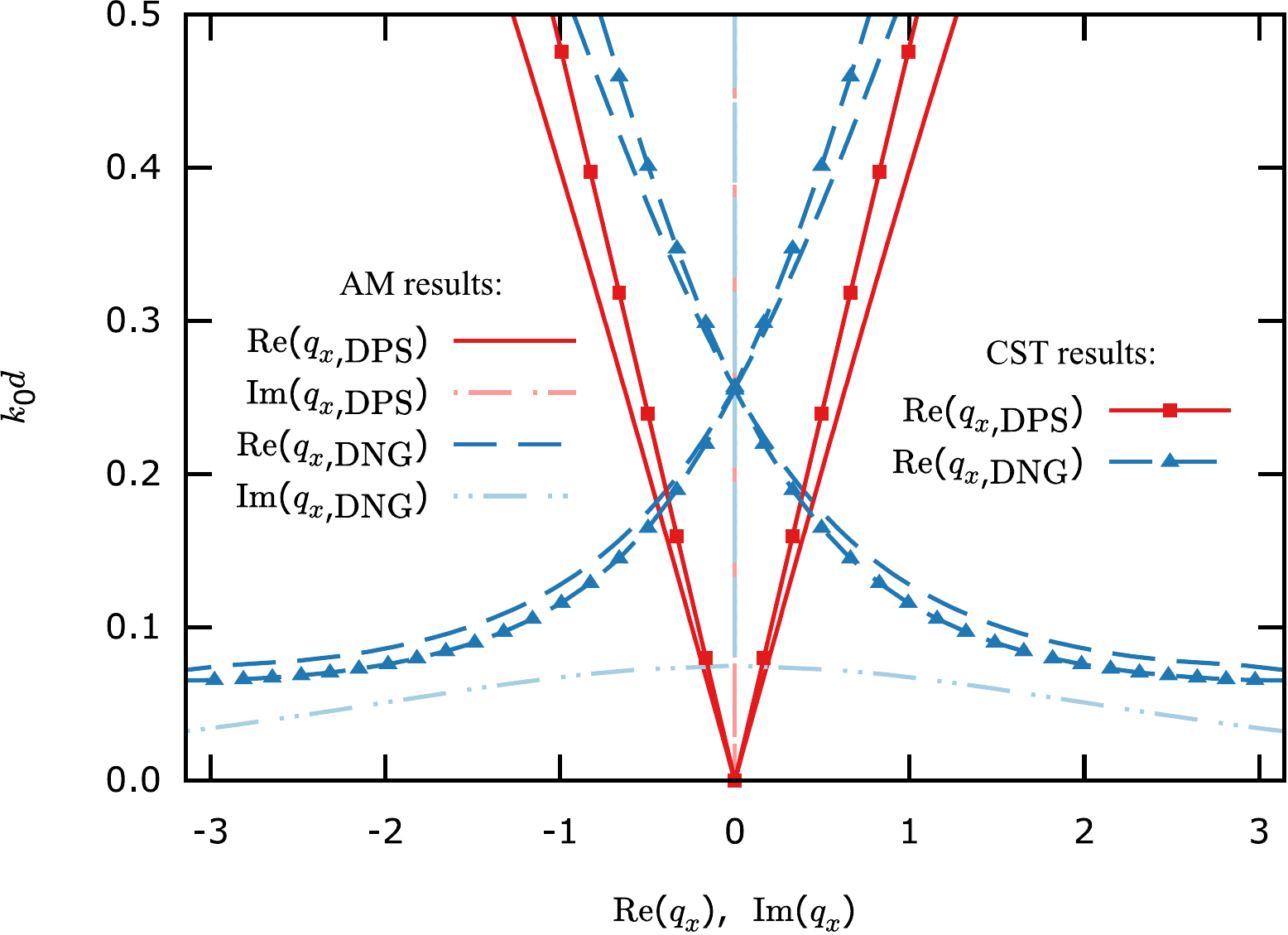}
  \caption{\label{dispBloch} (Color online) Dispersion of the Bloch
    waves in the uniform DNG and DPS networks realized by meshes of
    loaded strip lines obtained with the analytical model (AM) and
    with the eigenmode solver of the CST Microwave Studio (CST). The
    propagation direction is along the $x$-axis. Here, $k_0d$ is the
    normalized frequency: $k_0d = \omega d/c$ and $q_x$ is the
    normalized propagation factor: $q_x = k_xd$. The unit cell size is
    $d = 5$~mm. The characteristic impedance of the strip lines in
    both DNG and DPS networks is $Z_0 = 71.6$~Ohm. The relative
    permittivity of the dielectric is $\E_{\rm r} = 3$. In the
    AM, the load inductance and capacitance in the DNG
    network are $L = 5.16$~nH, and $C = 2.01$~pF, as found from
    Eqs.~\r{Zl0} and~\r{Yl0}, assuming that, at the operation frequency,
    $\beta_0d = \sqrt{\E_{\rm r}}k_0d = 0.315$. In the CST simulations
    with the unit cells from Fig.~\ref{stripcell}~(a,b), $L = 5.77$~nH
    and $C = 2.11$~pF.}
\end{figure}

The typical dispersion curves for the Bloch waves propagating in the
uniform conjugate-impedance matched DPS and DNG domains realized as
meshes of loaded strip lines are shown in Fig.~\ref{dispBloch}. These
results are obtained with the analytical model of Sec.~\ref{TLmesh}
(AM) and with unit cell simulations in the CST Microwave Studio
(CST). In Fig.~\ref{dispBloch}, the normalized free space wavenumber
$k_0d = (\omega d/c)$ is displayed on the vertical axis, and the
normalized Bloch wave propagation factor $q_x = k_xd$ is displayed on
the horizontal axis.

Let us first discuss the results of the AM. For this
case, the curves for both real and imaginary parts of $q_x$ are shown
in Fig.~\ref{dispBloch}.  In the considered range of frequencies, the
wave dispersion in the DPS network is similar to the dispersion of a
plane wave in a dielectric. Indeed, in this example, the relative
permittivity of the dielectric that fills the strip lines is
$\E_r = 3$, and, therefore, as follows from Eq.~\r{kxdps0}, at low
frequencies, $k_x = \sqrt{2\E_{\rm r}}k_0 \approx 2.45 k_0$. The wave
dispersion in the DNG network exhibits a band gap region at the
normalized frequencies $k_0d \lesssim 0.075$, where the normalized
Bloch wave propagation factor $q_x$ is such that $\Re\,q_x = \pm \pi$,
$\Im\,q_x \neq 0$. The region $0.075\lesssim k_0d \lesssim 0.256$ is
the backward wave propagation region, in which $q_x(\d q_x/\d\o) < 0$.
The dispersion curves for the waves in the DPS and the DNG networks
intersect in this range at $k_0d\approx 0.182$, $q_x\approx\pm 0.447$.

\begin{figure}[tb]
  \includegraphics[width=0.5\linewidth]{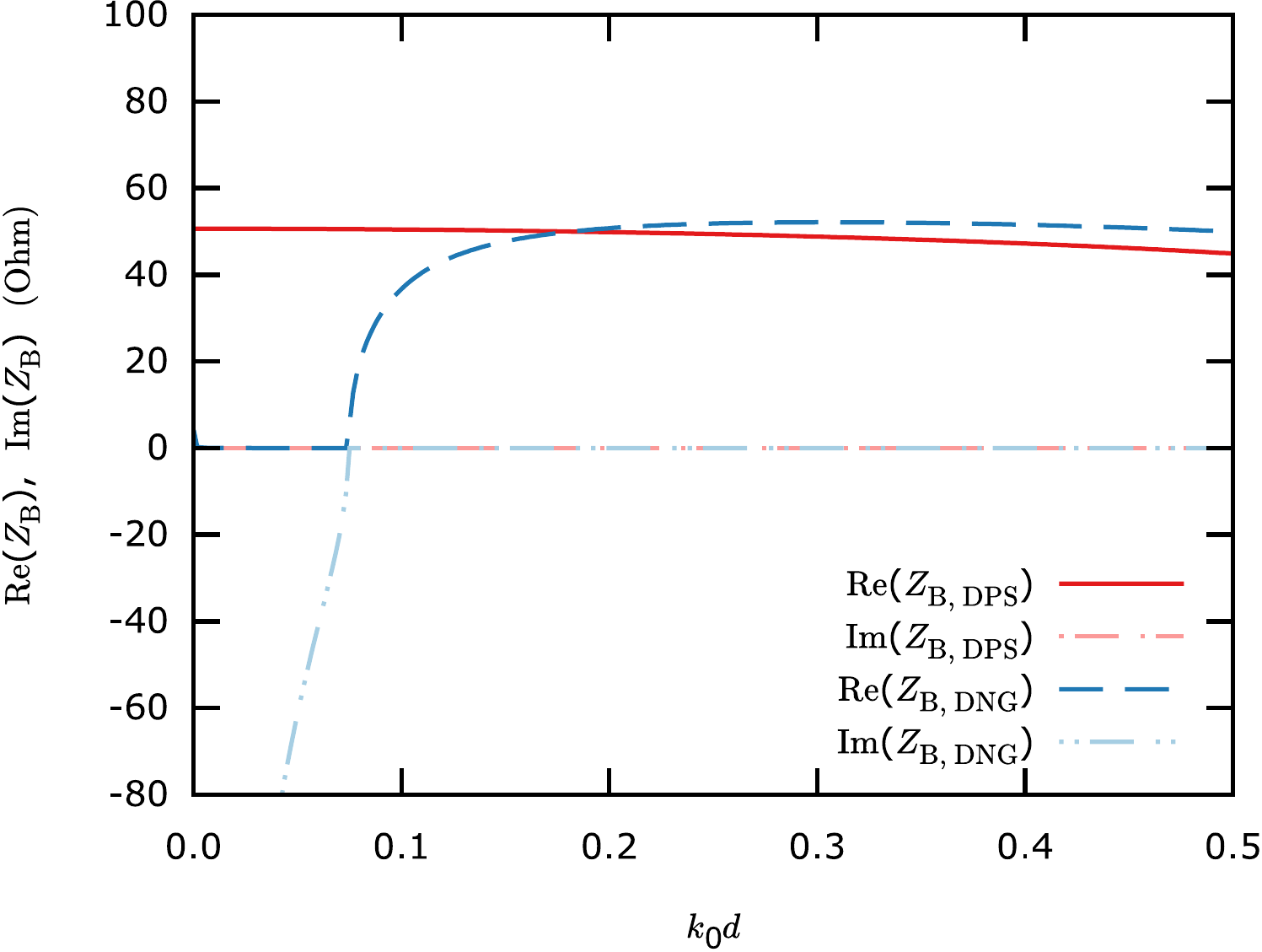}\\\hspace{12mm}(a)\\
  \includegraphics[width=0.5\linewidth]{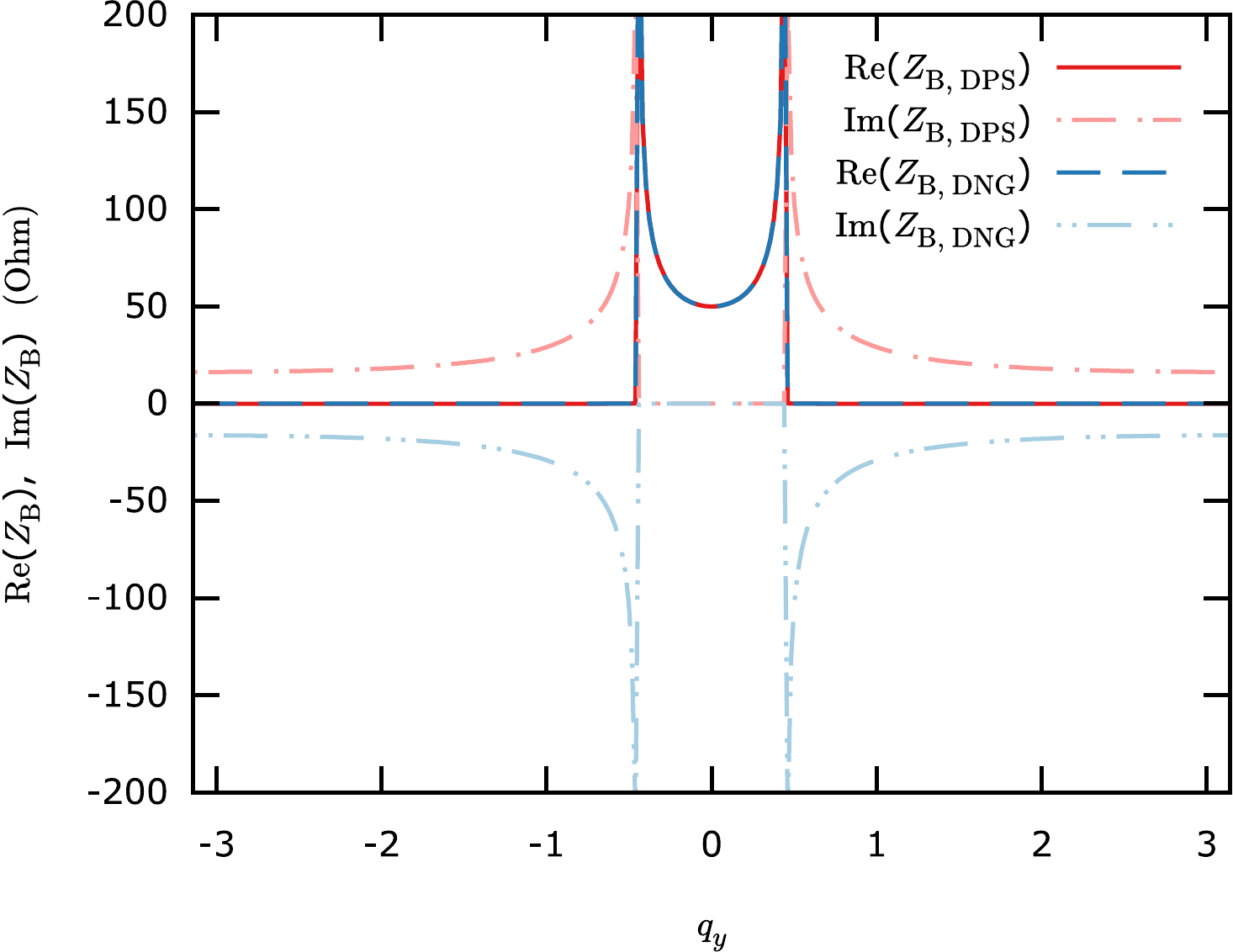}\\\hspace{12mm}(b)\\
  \caption{\label{impBloch} (Color online) Panel (a): Bloch wave
    impedances $Z_{\rm B}^{\rm DPS}$ and $Z_{\rm B}^{\rm DNG}$ for the
    waves propagating along the $x$-axis in both DPS and DNG networks
    as functions of the normalized frequency $k_0d$. Panel (b): The
    same impedances for the conjugate-impedance matched networks as
    functions of the normalized transverse wavenumber $q_y = k_yd$, at
    the normalized frequency $k_0d = 0.182$. The curves for
    $\Re\,Z_{\rm B}^{\rm DPS}$ and $\Re\,Z_{\rm B}^{\rm DNG}$ are
    indistinguishable from each other on the scale of this plot. The
    parameters used in both panels are the same as in
    Fig.~\ref{dispBloch} for the AM.}
\end{figure}

At $k_0d \approx 0.256$ the backward wave dispersion branch of the DNG
network transition into the forward wave dispersion branch of the same
network. Normally, a second band gap opens around this point. However,
this band gap closes when the characteristic impedances of the DPS and
the DNG TL segments match and the values of $L$ and $C$ loads are
obtained from the conjugate-impedance match conditions [Eqs.~\r{Zl0}
and~\r{Yl0}], which is our case.

The dependence of the Bloch wave impedance of the DPS and DNG domains
on the frequency (obtained with the AM) is shown in Fig.~\ref{impBloch}~(a). In the
considered range of frequencies, the wave impedance of the DPS network
is close to 50~Ohm. The wave impedance of the DNG network is purely
imaginary in the bandgap region $k_0d \lesssim 0.075$. Above this
region, $Z_{\rm B}^{\rm DNG}$ is real and positive. The curves for
$Z_{\rm B}^{\rm DPS}$ and $Z_{\rm B}^{\rm DNG}$ intersect at the point
$k_0d \approx 0.182$, which is the same as the intersection point of
the dispersion curves of the same networks. At this point
$Z_{\rm B}^{\rm DPS} = Z_{\rm B}^{\rm DNG} = 50$ Ohm. Moreover, a
direct numerical calculation shows that at this frequency point, the
equality $Z_{\rm B}^{\rm DNG} = \left(Z_{\rm B}^{\rm DPS}\right)^*$
holds for the Bloch waves with arbitrary transverse wavenumbers
$-\pi/d \le k_y \le \pi/d$, which confirms that the two domains are
perfectly conjugate-impedance matched. The dependency of
$Z_{\rm B}^{\rm DPS}$ and $Z_{\rm B}^{\rm DNG}$ on the transverse
wavenumber is depicted in Fig.~\ref{impBloch}~(b).

The dispersion of the same DPS and DNG cells has also been studied
with the eigenmode solver of the CST Microwave Studio. The geometry of
the DPS unit cell as realized in the CST Microwave Studio is shown in
Fig.~\ref{stripcell}~(a). Several DNG unit cell models were tested, in
which the load capacitance $C$ and the load inductance $L$ were
realized by different means. For instance, the capacitance $C$ was
realized with a gap in the strip line capped with a metallic patch (to
increase the gap capacitance), and the inductance $L$ was realized
with a thin helical wire. Realizations using standard surface mounted
(SMD) chip components were also checked. It has been found that
independently of the way how these loading elements are realized, the
modal dispersion resulting from the full wave simulations can be made
nearly coincident with the results of the AM (within the frequency
range $k_0d \lesssim 0.5$) by tuning the geometrical parameters of the
capacitors (the gap width, the interlacing area) and the inductors
(the wire radius, the number of turns). Therefore, in what follows, we
discuss the results for the unit cell structure shown in
Fig.~\ref{stripcell}~(b), which models the load inductors and
capacitors as effective impedance boundary conditions on an edge or a
line within the computational domain of the CST Microwave Studio, as
the most general representation of such loads.

The dispersion curves obtained with the CST Microwave Studio
eigenmode solver for the DNG and DPS cells with the geometries shown
in Fig.~\ref{stripcell}~(a,b) are depicted in Fig.~\ref{dispBloch}
alongside the AM results. Because the eigenmode solver
allows only for calculation of the dispersion of the propagating modes
(i.e., the evanescent modes are excluded) this figure only displays the
real part of the normalized propagation factor $q_x$ as a function of
the normalized free space wavenumber $k_0d$ for this
case. Note that
the values of the lumped loads ($L$ and $C$) differ slightly in the
CST calculations, as compared to the AM calculations. This is because
we have tuned these parameters in order to match the position of the
closed band gap in both cases. The observed residual discrepancy is
due to simplifications in the AM.

The CST eigenmode solver simulations were performed by considering up
to four propagating modes. Fig.~\ref{dispBloch} shows the dispersion
curves for the first mode of the DPS unit cell, and for the first and
the third mode of the DNG unit cell. It has to be noted that, for the
DNG cell, the CST Microwave Studio eigenmode solver predicts existence
of a spurious mode in a very narrow band close to
$k_0d \approx 0.256$, which is right within the closed bandgap between
the backward wave and the forward wave dispersion branches. This
spurious mode (the second mode as found by the eigenmode solver) is
related to a slight asymmetry in the transmission between the ports
$(x_{i-1},y_i)$ and $(x_{i+1},y_i)$ [or $(x_i,y_{i-1})$ and
$(x_i,y_{i+1})$] and the ports $(x_{i\pm 1},y_i)$ and
$(x_i,y_{i\pm 1})$ (see Fig.~\ref{2DTL}) in the real unit cell
structure depicted in Fig.~\ref{stripcell}~(b). This mode is not shown
in Fig.~\ref{dispBloch}.

\subsection{Numerical results of FDTLM simulations}

\begin{figure}[!htb]
  \centering
  \includegraphics[width=0.5\linewidth]{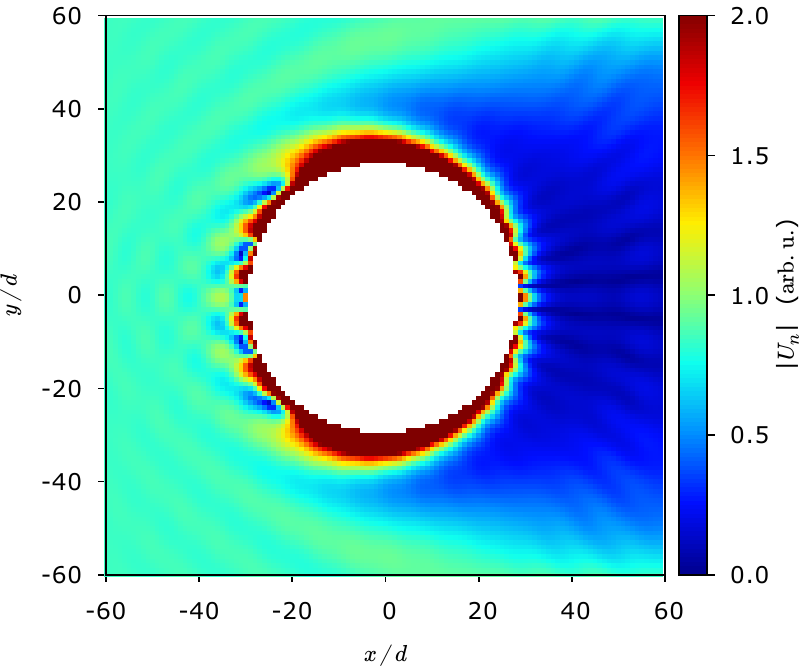}\\(a)\\
  \includegraphics[width=0.5\linewidth]{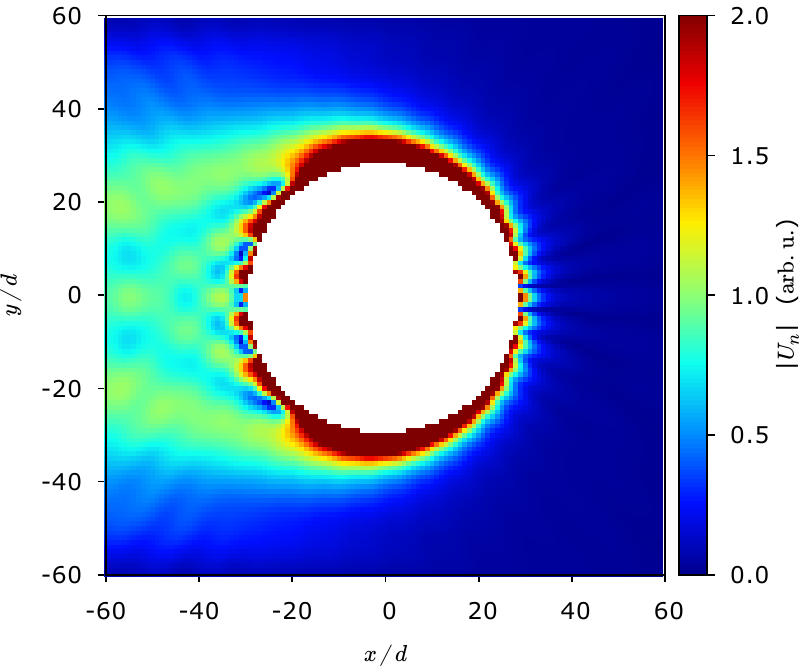}\\(b)\\
  \caption{\label{field_wh} (Color online) Distributions of the nodal
    voltage $|U_n|$, $n = \lfloor 120(60+y/d) + (60 + x/d)\rfloor+1$,
    $-60 \le x/d < 60$, $-60 \le y/d < 60$, for the wormhole structure
    under the plane wave incidence of unitary amplitude (see main text), in both the DPS (a) and the DNG (b) domains as
    functions of the normalized coordinates $x/d$ and $y/d$.
    The radius of the wormhole is
    $R_{\rm WH} = 30d$. The electrical parameters are:
    $\beta_0d = 0.315$, $\tan\delta = 10^{-4}$.}
\end{figure}

\begin{figure}[!htb]
	\centering
	\includegraphics[width=0.5\linewidth]{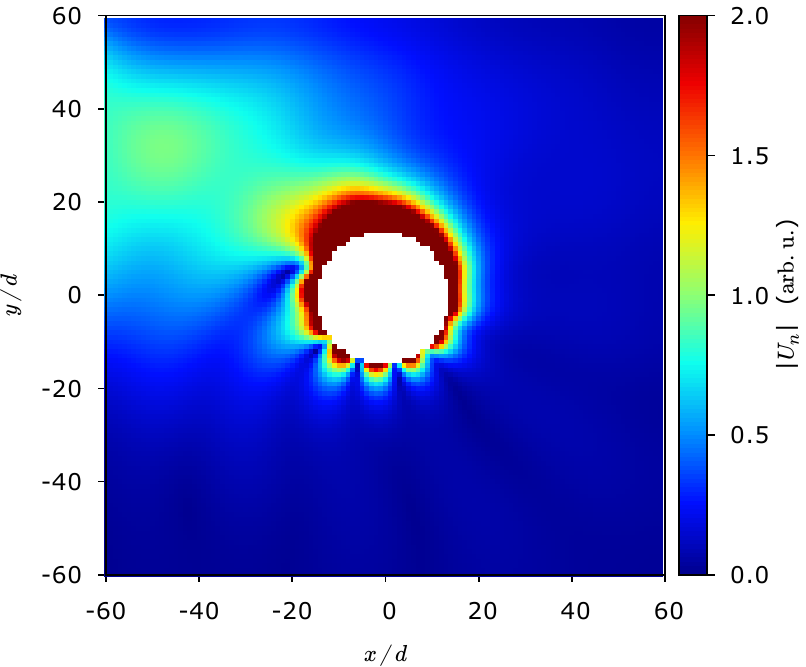}\\(a)\\
	\includegraphics[width=0.5\linewidth]{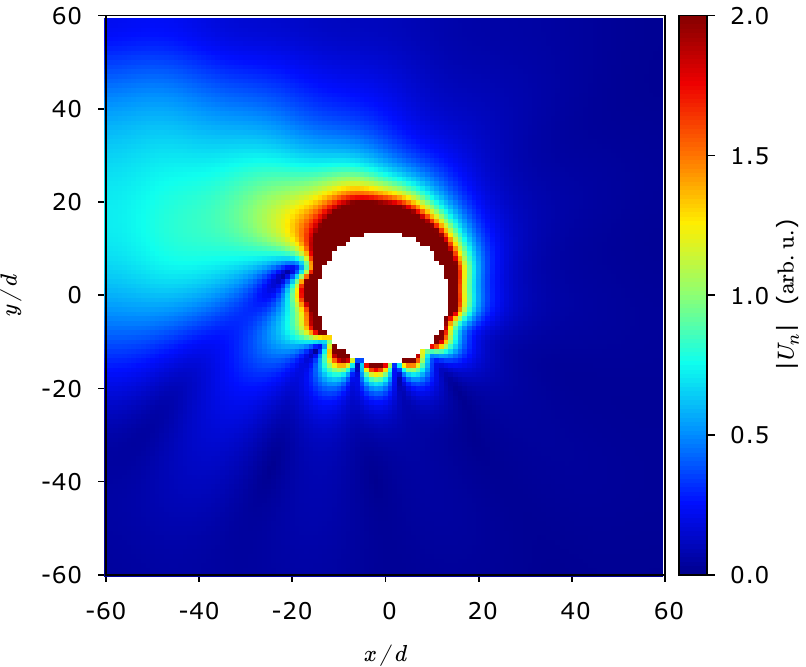}\\(b)\\
	\caption{\label{beam}(Color online) Trapping of a Gaussian
          beam by the metamaterial wormhole. Panel (a): The
          distribution of nodal voltage $|U_n|$ in the DPS
          plane. Panel (b): Same for the DNG plane. The radius of the
          wormhole is $R_{\rm WH} = 15d$. The Gaussian beam source
          (see Appendix~\ref{AppB}) is located in the DPS plane at
          $x = -60d$, with its maximum at $y_0 = 30d$ and the width
          $w=40d$. The electrical parameters are $\beta_0d = 0.1$,
          $\tan\delta=10^{-4}$.}
\end{figure}

The results of numerical simulations obtained with the FDTLM approach
are shown in Fig.~\ref{field_wh}. In this case, the wormhole structure
is formed by the DPS and DNG domains occupying an area of $120\x 120$
cells (with each cell being a square of size $d\x d$) and a wormhole
with the radius $R_{\rm WH} = 30 d$. The structure is excited by a
plane wave Huygens source (see Appendix~\ref{AppB} for details)
enclosing the whole DPS plane. Hence, Fig.~\ref{field_wh} depicts the
total (i.e.\ incident plus scattered) field. The source amplitude is
such that $\sqrt{1 - |\Gamma_0|^2}|V^{\rm inc}_0| = 1$ arb.~u., where
$V^{\rm inc}_0$ is the incident wave voltage (measured in arb.~u.)\
at the input ports of the unit cells located at $x = -60d$ and
$\Gamma_0$ is the Bloch wave reflection coefficient in these ports
(see Appendix~\ref{AppB}).

There are no cells at the middle of the domains where
$r < R_{\rm WH}$. Respectively, in this middle region (shown in white
in the figure) there is no propagation. Instead, an incident wave in
the DPS domain, when reaching the wormhole, passes to the DNG domain
through the connections at the wormhole edge. As is seen from
Fig.~\ref{field_wh}, after passing through the wormhole, the
transmitted wave forms a beam that propagates in the DNG domain in the
opposite direction with respect to the propagation direction of the
incident wave.

One can also see that a shadow is formed in the DPS region behind the
wormhole. Moreover, the diameter of the shadow is greater than the
diameter of the wormhole, which indicates that the effective absorption
cross section of the wormhole is such that
\[
  \sigma_{\rm norm} = {\sigma_{\rm abs}\over 2R_{\rm WH}} > 1,
\]
where $\sigma_{\rm norm}$ is the normalized absorption cross section
and $\sigma_{\rm abs}$ is given by Eq.~\r{sigmaabs} from
Appendix~\ref{AppB}. In this case, the numerically calculated value of
$\sigma_{\rm norm}$ is $1.46$, i.e., this metamaterial wormhole object
performs about 50\% better than the ideal black body
absorber. Note that this result is achieved for the object with a
rather large electrical size $\beta_0R_{\rm WH} = 30\beta_0d = 9.45$,
which means that the circumference of the object is on the order of 10
wavelength. Fig.~\ref{field_wh} also shows that there are practically
no reflections from the front of the wormhole in the DPS region.

Fig.~\ref{beam} illustrates trapping of nearby passing beams of
radiation by the metamaterial wormhole. In this example, we have
decreased the electrical cell size to $\beta_0d = 0.1$ and the
wormhole radius to $R_{\rm WH} = 15d$ in order to obtain a more
pronounced effect. When illuminated by a plane wave (not shown in
Fig.~\ref{beam}) the normalized absorption cross section of this
object is $\sigma_{\rm norm} \approx 2.7$, which means that the
object's shadow radius is about $2.7R_{\rm WH} \approx 40d$.

Since the maximum of the incident Gaussian beam is at $y_0 = 30d$, it
falls within the ``interception range'' of the metamaterial
superabsorber. We can see from Fig.~\ref{beam}~(a) that most of the
energy of the beam is captured by the wormhole. After passing through
the wormhole neck, the beam propagates in the DNG plane in the opposite
direction [Fig.~\ref{beam}~(b)] and its amplitude decreases (and the width increases) due to
diffraction, until the beam gets absorbed at the edge of the DNG
domain.

\begin{table}[!t]
	\centering
	\begin{tabular}{|c|c|c|c|c|c|c|c|c|c|c|c|c|c|c|c|}
		\hline
		$r/d$ & 30& 29& 28& 27& 26& 25& 24& 23\\ \hline
		$\beta_0d$ &
		0.315&
		0.397&
		0.284&
		0.340&
		0.719&
		0.576&
		0.623&
		0.777
		\\ \hline
		$Z_0/Z_0^{\rm DPS}$ &
		1&
		0.776&
		1.525&
		0.781&
		0.961&
		0.824&
		1.05&
		1.04
		\\ \hline
		$\sigma_{\rm norm}$ &
		--&
		1.05&
		1.22&
		1.13&
		1.21&
		1.22&
		1.20&
		1.15
		\\ \hline                          
	\end{tabular}\\
	\begin{tabular}{|c|c|c|c|c|c|c|c|c|c|c|c|c|c|c|c|}
		\hline
		$r/d$ & 22& 21& 20& 19& 18& 17& 16& 15\\ \hline
		$\beta_0d$ &
		0.757&
		0.675&
		0.609&
		1.076&
		0.674&
		1.313&
		0.921&
		0.808
		\\ \hline
		$Z_0/Z_0^{\rm DPS}$ &
		1.28&
		1.10&
		1.25&
		1.14&
		1.11&
		1.05&
		0.778&
		0.972
		\\ \hline
		$\sigma_{\rm norm}$ &
		1.21&
		1.18&
		1.14&
		1.16&
		1.18&
		1.07&
		1.09&
		1.10
		\\ \hline
	\end{tabular}
	\caption{\label{tab1}Parameters of the concentric rings of the DNG cells
		forming the object: The relative inner ring radius $r/d$, the
		normalized propagation factor parameter $\beta_0d$, the normalized
		characteristic impedance $Z_0/Z_0^{\rm DPS}$, and the normalized
		absorption cross section $\sigma_{\rm norm}$. The first column with
		$r=R_{\rm obj} = 30d$ lists the electrical parameters of the DPS
		plane.}
\end{table}

Although the results of Fig.~\ref{field_wh} and Fig.~\ref{beam}
confirm the presence of the superabsorption effect and illustrate the
main phenomena associated with it, from the application point of view,
it would be interesting to consider if the same effect could be
demonstrated in a structure in which the number of the DNG cells was
greatly reduced. Specifically, it is interesting to study the
situation in which the DNG cells (in both DPS and DNG domains) occur
only within a region of limited radius $r < R_{\rm obj}$, and with a
wormhole of the radius $R_{\rm WH} < R_{\rm obj}$.

It is immediately understood that, in this case, the electrical
parameters of the DNG cells in the DPS domain must vary with $r$. From
the theoretical results for continuous media (Sec.~\ref{summary}, the
cylindrical case), one would expect that the normalized propagation
factor $\beta_0d$ should vary with radius as
$\beta_0d \propto \sqrt{\E_{zz}\M_{\varphi}} \propto r^{-2}$ and the
line impedance as
$Z_0 \propto \sqrt{\M_{\varphi}/\E_{zz}} \propto r^2$. However, direct
numerical FDTLM calculations show that, for cells with reasonably small
(i.e., not too small) electrical thickness $\beta_0d \gtrsim 0.1$, the
results obtained when using the profiles deduced from the continuous
medium theory are far from being optimal. In fact, if
$\beta_0d \approx 0.3$ (which is an attainable value from the
practical point of view) in the uniform DPS region and at the object's
border at $r = R_{\rm obj}$, then inside the object at, for example,
$r = R_{\rm obj}/2$, the same parameter must reach the value of
$4\x0.3 = 1.2$, which is already too large for the cell to be
considered electrically small, and thus the continuous medium
approximation fails.

\begin{figure}[t]
	\centering
	\includegraphics[width=0.5\linewidth]{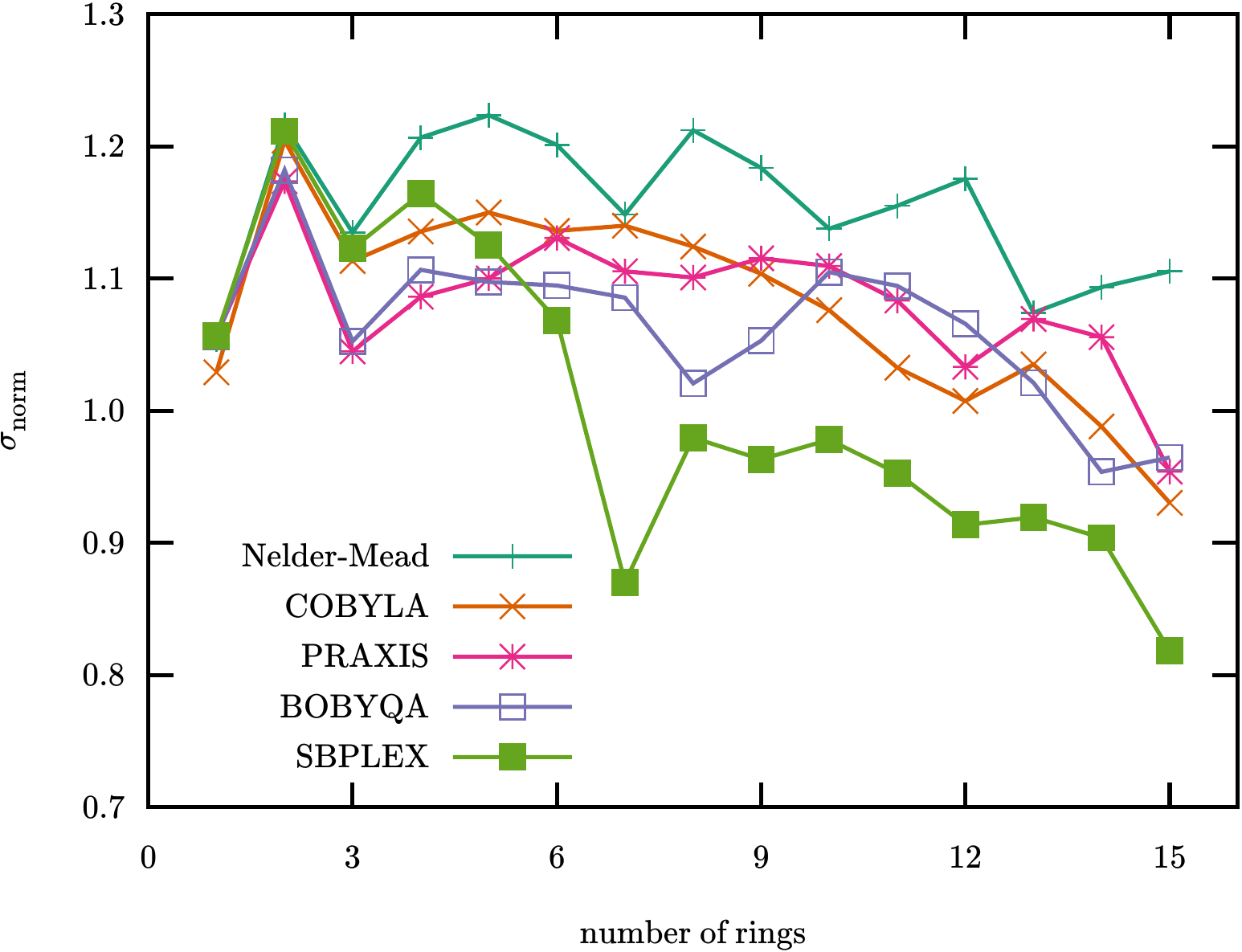}
	\caption{\label{sigma} (Color online) The normalized absorption cross section
		$\sigma_{\rm norm}$ of the object formed by a number of concentric
		rings of the DNG cells with varying parameters as a function of the
		number of the object's rings. The radii of the rings are as in
		Tab.~\ref{tab1}. The electrical parameters of each ring are
		obtained by using a number of numerical optimization algorithms:
		Nelder-Mead, COBYLA, PRAXIS, BOBYQA, and SBPLEX.~\cite{nlopt}}
\end{figure}

Therefore, in the following numerical simulations, an optimization
approach for such a structure is used, the purpose of which is to
establish some optimal variation profiles for $\beta_0d$ and $Z_0$
within the object. We consider an object with the outer diameter
$R_{\rm obj} = 30d$, which is formed by up to 15 concentric rings of
the DNG cells. The radii of these rings and the electrical parameters of
the cells in these rings (as obtained by an optimization procedure)
are listed in Tab.~\ref{tab1}. The table lists only the inner
radii. The outer radius of a given ring is the inner radius of the
previous ring, and so on, until the ring of the smallest radius is
reached. The inner radius of the smallest ring is
$r = R_{\rm WH} = 15d$, at which point the wormhole starts. In the DNG
domain the region $R_{\rm WH}<r<R_{\rm obj}$ is filled with the cells
whose electric parameters match to the parameters of the last ring.

\begin{figure}[!htb]
	\centering
	\includegraphics[width=0.5\linewidth]{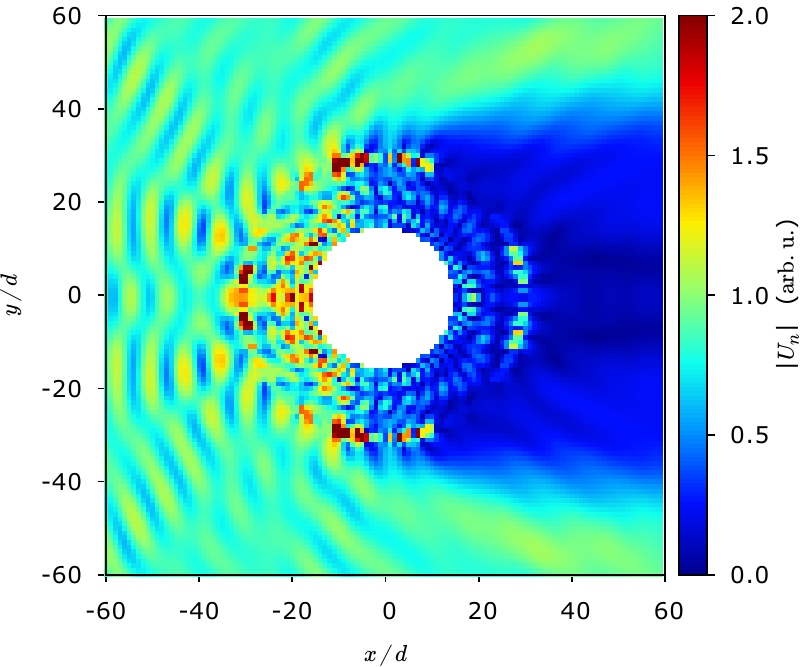}\\(a)\\
	\includegraphics[width=0.5\linewidth]{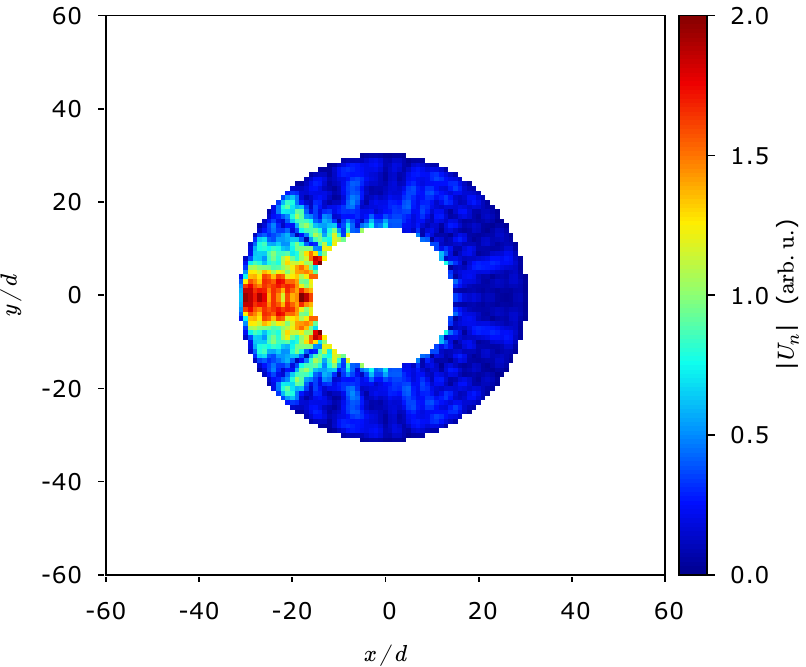}\\(b)\\
	\caption{\label{field_obj} (Color online) Distribution of the nodal voltage $|U_n|$
		(defined in the same way as in Fig.~\ref{field_wh}) for an object
		composed by a set of concentric rings of the DNG cells of varying
		effective $\beta_0^{\rm DPS}d$ and impedance $Z_0^{\rm DPS}$
		placed in the middle of the DPS domain [panel (a)] at
		$R_{\rm WH}<r<R_{\rm obj}$, whose inner ring is connected through
		the wormhole to the DNG domain [panel (b)] formed by a single
		uniform ring $R_{\rm WH}<r<R_{\rm obj}$ of the DNG cells. The
		structure is under the plane wave incidence of unitary
		amplitude. The object radius is $R_{\rm obj} = 30d$ and the
		wormhole radius is $R_{\rm WH} = 15d$. The electrical parameters
		in the DPS region where $r > R_{\rm obj}$ are $\beta_0d = 0.315$,
		and $\tan\delta = 10^{-4}$. The other parameters are listed in
		Tab.~\ref{tab1}.}
\end{figure}

\begin{figure}[!htb]
	\centering
	\includegraphics[width=0.5\linewidth]{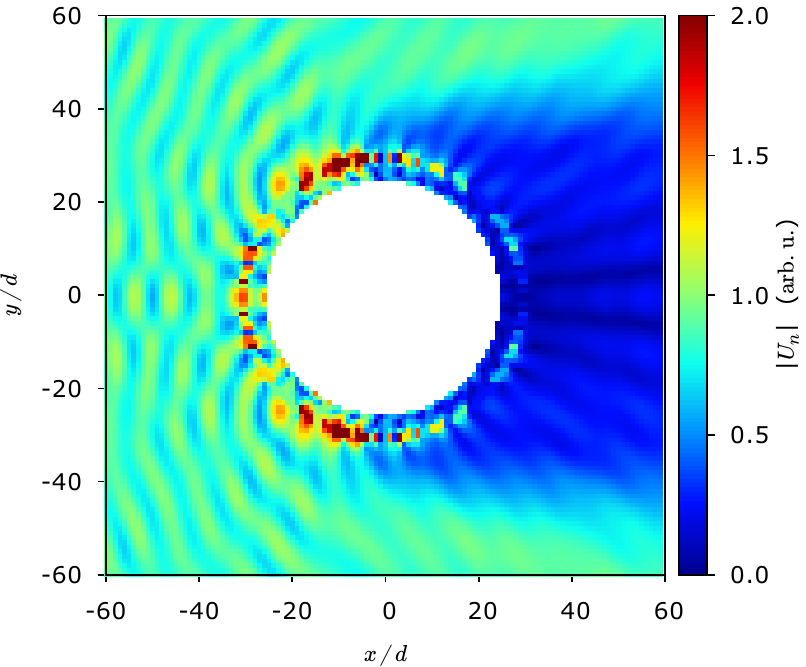}\\(a)\\
	\includegraphics[width=0.5\linewidth]{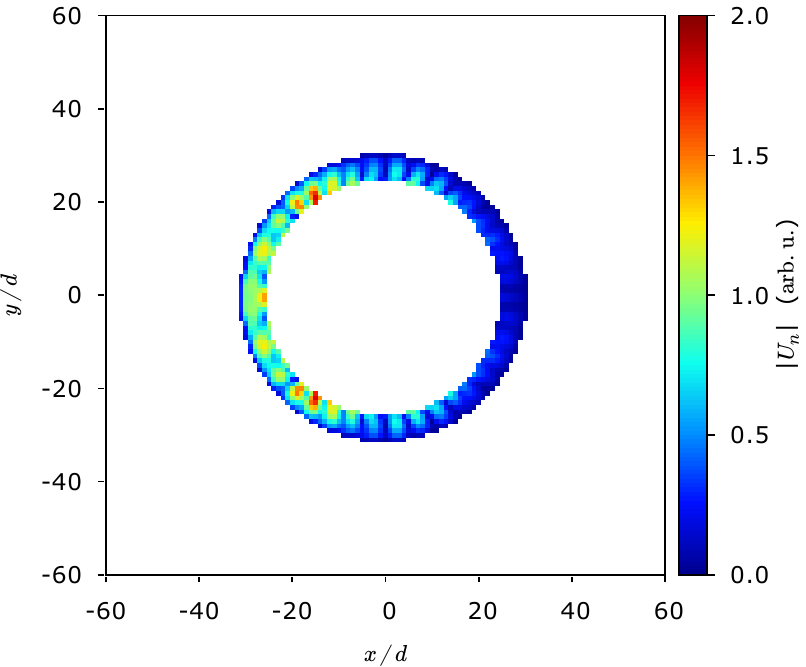}\\(b)\\
	\caption{\label{tworing} (Color online) Distribution of the nodal voltage $|U_n|$
		for an object composed by 5 concentric rings of the DNG cells in
		the DPS domain [panel (a)] and a single uniform ring $R_{\rm WH}<r<$
		of the DNG cells in the DNG domain [panel (b)]. The structure is
		under the plane wave incidence of unitary amplitude. The ring radii and the electrical
		parameters of the DNG cells are listed in Tab.~\ref{tab1}.}
\end{figure}

The electrical parameters of the rings are optimized in order to
maximize $\sigma_{\rm abs}$ of the whole object. The optimization is
performed in gradual steps. First, a structure containing only a
single ring with the largest radius is considered, with the wormhole
starting at the inner edge of this ring. Because there is only one
ring in the structure at this step, it is also the last ring, and the
parameters of the cells in the DNG domain are matched to this
ring. After the electrical parameters of this ring are optimized to
maximize $\sigma_{\rm abs}$, they are fixed, and the second ring is
added to the structure (which becomes the new last ring) and the
optimization of the electrical parameters of this ring is performed
under the same goal. In this way, the parameters of all 15 rings are
found. The optimization results are summarized in Fig.~\ref{sigma}.

The FDTLM simulation results for two ring structures under plane wave
incidence are shown in Figs.~\ref{field_obj} and~\ref{tworing}. As in
Fig.~\ref{field_wh}, there are no cells in the white areas. As
compared to the simple wormhole structure, the results of
Fig.~\ref{field_obj} show that this object has a much higher level of
reflections, which means that some part of the incident power is lost
due to these reflections, and, therefore, one may not expect to get a
high effective $\sigma_{\rm abs}$ for this object when a large number
of rings is used.

In fact, further numerical simulations demonstrate (Fig.~\ref{sigma})
that there is an optimal number of rings $n$, $1 < n < 15$, at which
$\sigma_{\rm abs}$ attains a maximum. Fig.~\ref{sigma} shows the
variation of the normalized absorption cross section,
$\sigma_{\rm norm} = \sigma_{\rm abs}/(2R_{\rm obj})$, as a function of
the object's rings number, as in the gradual optimization procedure
described earlier. For the selected value of the $\beta_0d$ parameter
in the DPS region: $\beta_0d = 0.315$, the optimal ring number is
$n=5$, with the rings' radii and the electrical parameters listed in the
beginning of Tab.~\ref{tab1} (these parameters are obtained with the
Nelder-Mead optimization algorithm from NLopt
library~\cite{nlopt}). The normalized absorption cross section in this
case is $\sigma_{\rm norm}=1.224$.

The distribution of the field around this object with a rather small
number of the DNG cells is shown in Fig.~\ref{tworing}. It is
interesting that the performance of this object is still more than
20\% higher than that of the black body absorber of the same diameter,
and such an increase in the performance is achieved with just a small
number of the DNG cells distributed around the perimeter of the
wormhole, as compared with the case of Fig.~\ref{field_wh}, where a
large number of the DNG cells distributed over the whole DNG domain is
used. We expect that with an even better realization strategy that
involves DNG cells of varying geometry and with better optimization
procedures these results can be further improved. In fact, preliminary
results obtained with a global optimization approach replacing the
ring-by-ring approach (to be reported elsewhere) indicate that in the
same setup one can achieve, at least, $\sigma_{\rm norm} \approx 1.3$.

\section{\label{concl}Conclusions}

In this work, possible realizations of the superabsorbing metamaterial
objects whose effective absorption cross section is significantly
greater than the geometrical cross section have been studied
theoretically and simulated numerically. The superabsorption effect
has been modeled with the metamaterial TL-based structures that support
effectively two-dimensional propagation of the electromagnetic
waves. We have shown that in this model, a finite-size
conjugate-impedance matched superabsorbing object can be equivalently
represented with a wormhole structure formed by two electrically
connected DNG and DPS domains. With respect to the waves propagating
within the DPS domain, the wormhole appears as a conjugate-impedance
matched absorber, while for the waves propagating in the DNG domain,
the wormhole acts as a radiation source. The waves transmitted to the
DNG domain through the wormhole neck are then absorbed at the edges of
the DNG domain, where the open ports are terminated with matched
loads.

By using the TL-based unit cells with the electrical size of
$\beta_0d = 0.1$ and the loss tangent value of $\tan\delta=10^{-4}$,
we have obtained the normalized absorption cross section about three
times greater than that for the black body of the same size. We have
demonstrated the trapping of nearby passing beams of radiation by
the metamaterial superabsorbers that has been predicted
earlier.~\cite{meta2016} For a larger wormhole structure with
$\beta_0d = 0.315$ and $\beta_0R_{\rm WH}=9.45$, the obtained
normalized absorption cross section is $\sigma_{\rm norm} = 1.46$,
which means that, even for objects with a relatively large
circumference of about 10 wavelengths, the metamaterial wormhole
superabsorber can outperform the black body absorber by about 50\%.

We have found that the superabsorption effect can be also observed in
non-uniform structures with a smaller number of DNG cells, as compared
to the complete wormhole structure. Namely, such an effect can be
observed for a DNG metamaterial object that fits entirely within a
region of a finite radius $r = R_{\rm obj}$, such that
$\beta_0R_{\rm obj} \gg 1$. This is especially interesting for
applications of the metamaterial superabsorbers as efficient
harvesters of electromagnetic radiation which absorb more energy from
an incoming plane wave than what is incident directly on their
surface.

It is worth noting that the wormhole structure can be also used to
demonstrate the narrow-band super-Planckian emitting property of the
conjugate-impedance matched superabsorbers.\cite{mmsuper} In order to
do this in practice, one will have to perform electric
(Johnson-Nyquist) noise measurements in this structure under controlled thermal
conditions. We reserve such a study for a future work.

\section*{Acknowledgment}
The authors acknowledge support under the project
Ref.~UID/EEA/50008/2013, sub-project MMSUPER, financed by Funda\c{c}\~{a}o
para a Ci\^{e}ncia e a Tecnologia (FCT)/Minist\'{e}rio da Ci\^{e}ncia,
Tecnologia e Ensino Superior (MCTES), Portugal. S.I.M. acknowledges
support from Funda\c{c}\~{a}o para a Ci\^{e}ncia e a Tecnologia (FCT), Portugal,
under Investigador FCT~(2012) grant
(Ref.\ IF/01740/2012/CP0166/CT0002).

\appendix

\section{\label{AppA}Scattering matrices of conjugate-impedance matched DPS and
  DNG cells}

The $S$-matrix of the square unit cell with the schematic depicted in
Fig.~\ref{2DTL} is a $4\x 4$ matrix whose elements, $S_{mn}$, satisfy
$S_{mn} = S_{nm}$ due to the reciprocity, and
$S_{11}=S_{22}=S_{33}=S_{44}$ due to the symmetry of the unit cell. In
addition, if the ports are numbered around the perimeter of the unit
cell (for example, in the counterclockwise direction), the unit cell
symmetry demands that $S_{12}=S_{14}$, $S_{13}=S_{24}$.

Note that because in the considered analytical model all four ports of
the unit cell are equivalent to each other, the $S$-parameter matrix
resulting from this model also must satisfy $S_{12} = S_{13}$, and
thus all elements of the $S$-matrix can be expressed through just a
pair of reflection and transmission coefficients ${\cal R}$ and
${\cal T}$ of an isolated unit cell, assuming that
its four ports are connected to infinite transmission lines with
characteristic impedance $Z_0$. The analytical expressions for
${\cal R}$ and ${\cal T}$ are
\begin{widetext}
\[
  {\cal R} = -{4(\bar{Y}+2\bar{Z}+2)\tan^2\!{\beta_0d\over
      2}-4j\bar{Z}(\bar{Z}+\bar{Y}) \tan{\beta_0d\over
      2}-\bar{Y}\bar{Z}^2-8\bar{Z}+4\bar{Y}+8\over
    \left(2j\tan{\beta_0d\over
        2}+\bar{Z}+2\right)\left(2j(\bar{Y}+2(\bar{Z}+2))\tan{\beta_0d\over
        2} +\bar{Y}\bar{Z}+2\bar{Y}+8\right)},
  \l{smatrref}
\]
and
\[
  {\cal T} = {8\left(\tan^2\!{\beta_0d\over 2}+1\right)\over
    \left(2j\tan{\beta_0d\over
        2}+\bar{Z}+2\right)\left(2j(\bar{Y}+2(\bar{Z}+2))\tan{\beta_0d\over
        2} +\bar{Y}\bar{Z}+2\bar{Y}+8\right)},
  \l{smatrtr}
\]
\end{widetext}
with $\bar{Z} = Z/Z_0$, and $\bar{Y} = Z_0Y$.

Respectively, the $S$-matrix elements of the unit cell are expressed
as
\[
  S_{mm} = {\cal R}, \quad S_{mn} = {\cal T}, \quad \forall\,m,n: m\neq n.
\]

When $\bar{Z} = \bar{Y} = 0$, which is the case of the DPS unit cell,
Eqs.~\r{smatrref} and~\r{smatrtr} reduce to
\[
  {\cal R}^{\rm DPS} = -{1\over 2}e^{-j\beta_0 d}, \quad {\cal T}^{\rm DPS} = {1\over 2}e^{-j\beta_0 d},
\]
which is the well-known result for the reflection and transmission
coefficients at a junction of four equal TL segments.\cite{Christopoulos}

In addition, it is easy to verify that when the values of the loads
$Z$ and $Y$ are taken in accordance with Eqs.~\r{Zl0} and~\r{Yl0}
for the conjugate-impedance matched DNG network with negligible loss
($\Im\,\beta_0\rightarrow 0$) and the real-valued characteristic TL impedance
$Z_0$, the reflection and transmission coefficients~\r{smatrref}
and~\r{smatrtr} reduce to
\[
  {\cal R}^{\rm DNG} = -{1\over 2}e^{+j\beta_0 d},\quad
  {\cal T}^{\rm DNG} = {1\over 2}e^{+j\beta_0 d},
\]
and thus the $S$-matrix of the conjugate-impedance matched DNG cell
(with the same real TL impedance $Z_0$) satisfies
\[
  S_{mn}^{\rm DNG} = \left(S_{mn}^{\rm DPS}\right)^*.
\]
This result can be also obtained directly from the relation between
the impedance matrix $\_Z$ (the $Z$-matrix) and the scattering matrix
$\_S$ of a single cell through the following expressions:
$\_S = (\_Z+Z_0\_I)^{-1}\.(\_Z-Z_0\_I)$ and
$\_Z = Z_0(\_I+\_S)\.(\_I-\_S)^{-1}$, where $\_I$ is the unit $4\x4$
matrix, and the generalized conjugate-impedance match
condition:\cite{SPIE}
$\_Z^{\rm DNG} = \left(\_Z^{\rm DPS}\right)^\dagger$ (here, the symbol
$^\dagger$ denotes the Hermitian conjugate).

\section{\label{AppB} Implementation of FDTLM method}

Given the $S$-matrices of all DPS and DNG cells, the wave propagation
in the whole structure can be studied by solving a system of linear
equations for the unknown wave amplitudes in the ports of adjacent
unit cells. Namely, if the DPS cells are numbered as $1\ldots N$ and
the DNG cells are numbered as $1\ldots M$, the unknown complex
amplitudes of the incident and reflected waves (which can be
understood as the waves of electric voltage) in all ports of all cells
can be collected into column vectors
$\_V^{\rm inc} = (a_1^{\rm DPS},\ldots, a^{\rm DPS}_{4N},a_1^{\rm
  DNG},\ldots, a^{\rm DNG}_{4M})^T$, and
$\_V^{\rm ref} = (b_1^{\rm DPS},\ldots, b^{\rm DPS}_{4N},b_1^{\rm
  DNG},\ldots, b^{\rm DNG}_{4M})^T$, respectively. The length of these
vectors is $4(N+M)$ due to the fact that each cell has four ports,
and, therefore, these vectors are composed of $N+M$ groups of four wave
amplitudes belonging to each cell.

As is evident from the above definition, the vectors $\_V^{\rm ref}$
and $\_V^{\rm inc}$ satisfy
\[
  \_V^{\rm ref} = \_S\.\_V^{\rm inc},
  \l{bigS}
\]
where
$\_S = \mbox{diag}(\_S_1^{\rm DPS},\ldots\_S_N^{\rm DPS},\_S_1^{\rm
  DNG},\ldots\_S_M^{\rm DNG})$ is a block-diagonal matrix formed by
the $4\x4$ scattering matrices of all the DPS and DNG cells in the wormhole
structure.

On the other hand, because the adjacent ports in the neighboring cells
are connected, the incident wave in one of such ports is essentially
the reflected wave in the other and {\it vice versa}. Therefore,
\[
  \_V^{\rm inc} = \_C\.\_V^{\rm ref},
  \l{bigC}
\]
where $\_C$ is the so-called connection matrix. The elements of $\_C$
are mostly zeros, with some elements $C_{mn}=1$, where $m$ and $n$ are
such that the electric connection of the respective ports in a pair of
adjacent cells demands that $V^{\rm inc}_m=V^{\rm ref}_n$. It is
evident that $C_{mn} = C_{nm}$. The general structure of the
connection matrix can be represented in the block matrix form as
\[
  \_C = \matr{\left[\_C_{\rm DPS}\right]_{(4N\x 4N)} & \left[\,\_C_{\rm WH}\,\right]_{(4N\x 4M)}\\[2mm]
    \left[\_C_{\rm WH}^T\,\right]_{(4M\x 4N)} & \left[\_C_{\rm DNG}\right]_{(4M\x 4M)}}.
\]

Note that besides connections between the DPS cells within the DPS
domain represented by $\_C_{\rm DPS}$, and similar connections between
the DNG cells in the DNG domain represented by $\_C_{\rm DNG}$, there
are also connections between the DPS and DNG cells at the wormhole
neck, which are taken into account by the off-diagonal blocks
$\_C_{\rm WH}$ and $\_C_{\rm WH}^T$.

Additionally, the diagonal elements $C_{mm}$ with indices $m$ that
correspond to unconnected ports at the edges of the structure can be
set to a non-zero value in order to realize an absorbing boundary
condition (ABC) which will imitate infinite continuation of the
periodic DPS or DNG mesh. A good approximation for such ABC at the
edge of the DPS or the DNG structure is $C_{mm} = \Gamma_0$, where
\[
  \Gamma_0 = {Z_\perp-Z_0\over
  Z_\perp+Z_0},
\]
with $Z_\perp$ being the Bloch impedance for the wave in the DPS (or
the DNG) mesh that impinges normally at the interface where the ABC is
defined.

When external sources are present, Eq.~\r{bigC} must be modified in
order to include the contribution to the incident waves due to such
sources:
\[
  \_V^{\rm inc} = \_C\.\_V^{\rm ref} + \_V^{\rm ext}.
  \l{Csrc}
\]
By combining Eq.~\r{bigS} with Eq.~\r{Csrc} we find that
\[
  (\_I - \_C\.\_S)\.\_V^{\rm inc} = \_V^{\rm ext}.
  \l{bigsys}
\]
In the simulations of the wormhole structure, Eq.~\r{bigsys} is
solved for $\_V^{\rm inc}$ for a given excitation vector
$\_V^{\rm ext}$. Next, the vector $\_V^{\rm ref}$ is found from
Eq.~\r{bigS}. The vector of total voltages at the input ports of all unit cells is
then obtained as $\_V = \_V^{\rm inc} + \_V^{\rm ref}$, from which we
can express the electric voltage at the middle point of $n$-th DPS
unit cell, by using the equivalent network of Fig.~\ref{2DTL} (with
$Z=Y=0$):
\[
  U_n^{\rm DPS} = {\ds\sum\limits_{i=1}^4V_{4n-4+i}\over \ds4\cos{\beta_0 d\over 2}}, \quad 1 \le n \le N,
\]
and, at $m$-th DNG cell, as
\begin{align}
  U_m^{\rm DNG} =
               {\ds\sum\limits_{i=1}^4V_{N+4m-4+i}\over
  \ds\left(4+{YZ\over 2}\right)\cos{\beta_0 d\over 2} +
  j\left(YZ_0+{2Z\over Z_0}\right)\sin{\beta_0d\over 2}},\quad 1 \le m \le M,
\end{align}
which reduces to 
\[
  U_m^{\rm DNG} = {\ds\sum\limits_{i=1}^4V_{N+4m-4+i}\over \ds4\cos{\beta_0 d\over 2}}, \quad 1 \le m \le M,
  \l{Unode}
\]
when the load impedance $Z$ and the load admittance $Y$ are given by
Eqs.~\r{Zl0} and~\r{Yl0}.

Eqs.~\r{bigS}--\r{Unode} constitute the main theoretical formulation
of the employed FDTLM method. Note that the external source vector
$\_V^{\rm ext}$ can be initialized in many different ways, allowing
for a number of excitation scenarios to be studied. For instance, the
excitation by an incident 2D Gaussian beam can be modeled by setting
up the elements $V^{\rm ext}_n$ that correspond to the open ports at
one of the four edges of the DPS domain (for instance, at
$x = -N_xd/2$, where $N_x$ is the number of cells across the whole
structure along the $x$-axis, with $x=y=0$ being at the middle of the
DPS domain) to values proportional to
$e^{-jk_{\rm t} y-(y-y_0)^2/w^2}$, with $y_0$, $w$, and $k_{\rm t}$
being the parameters of the beam, and $y$ being the coordinate along
the edge. Such a source will produce a Gaussian beam that propagates
in the DPS domain in the direction defined by the wave vector
$\_k = (k_x,k_y)$, where $k_y = k_{\rm t}$ and $k_x$ is given by
Eq.~\r{kxdps0}. The parameter $w$ defines the width of the beam, and
the parameter $y_0$ sets the initial location of its maximum.

The external source can also be defined for an effective plane wave
excitation scenario. Moreover, in this case it is possible to set up
the source in a way that mimics electromagnetic excitation of a body
by equivalent Huygens sources --- pairs of orthogonal electric and
magnetic dipole moments defined on a surface fully enclosing the
object under study.\cite{Lindell} Because the field of such sources
vanishes outside the enclosed domain, the outside field is just the
field scattered by the body, i.e., excitation of an object by such
sources allows for a straightforward determination of the field
scattered by the object.

In an FDTLM simulation, such a Huygens source can be constructed in
the following way. First, based on the results of the analytical model
of Sec.~\ref{TLmesh}, a vector $\_V^{\rm inc} = \_V^{\rm DPS}_{\rm B}$
that corresponds to a Bloch wave solution in a uniform DPS domain with
no scatterers (i.e., without the wormhole or any other irregularities)
is formed. Second, a mask $\_M = \mbox{diag}(M_n)$, $M_n\in \{0,1\}$,
is applied by calculating the product $\_M\.\_V^{\rm DPS}_{\rm B}$
such that it filters out the elements that relate to the unit cells
outside the domain that we wish to be enclosed by the Huygens
source. Finally, the Huygens source vector $\_V^{\rm ext}_{\rm H}$
that creates the plane-wave-like incident field inside the enclosed
domain and the zero field outside is found as
\[
  \_V^{\rm ext}_{\rm H} = (\_I - \_C^{\rm DPS}_0\.\_S^{\rm DPS}_0)\.\_M\.\_V^{\rm DPS}_{\rm B},
  \l{tlmHuyg}
\]
where $\_C^{\rm DPS}_0$ and $\_S^{\rm DPS}_0$ are the connection and
scattering matrices of the uniform, unperturbed DPS domain. In order
to determine the wormhole behavior under such excitation, the
source~\r{tlmHuyg} is substituted into Eq.~\r{bigsys} and the
resulting matrix system is solved for $\_V^{\rm inc}$. The obtained
solution will relate to the total (incident plus scattered) field in
the DPS region enclosed by the Huygens source and to the transmitted
field in the DNG region, while in the DPS region outside the enclosed
domain, it will relate only to the scattered field.

Finally, the total electric power that enters into the wormhole under
a given excitation and becomes eventually absorbed due to the ABCs at
the edges of the DNG domain can be expressed as (we understand the
time-harmonic voltages as rms values)
\[
  P_{\rm abs} = \Re\left[{1\over Z_0}{\_V^{\rm inc}}^\dagger\.
  \matr{\_0 & -\_C_{\rm WH}\\ \_C_{\rm WH}^T & \_0}\.\_V^{\rm ref}\right],
\]
and the effective absorption cross section $\sigma_{\rm abs}$ of the
wormhole is calculated as
\[
  \l{sigmaabs}
  \sigma_{\rm abs} = {P_{\rm abs}\over \Pi_{\rm inc}},
\]
where $\Pi_{\rm inc}$ is the density of the incident power flux, which
is determined by the amplitude of the incident wave. For example, for
an incident Bloch wave propagating in the DPS domain along the
$x$-axis and characterized by the incident wave amplitude
$V^{\rm inc}_0$ in the input ports at the edge $x = -N_x d/2$, where
the wave enters the structure,
\[
  \Pi_{\rm inc} = {1\over Z_0d}(1-|\Gamma_0|^2)|V^{\rm inc}_0|^2.
\]
Note that in the 2D scattering problem that we consider,
$\sigma_{\rm abs}$ has the dimension of a length (and the physical
meaning of a characteristic diameter) rather than an area (as in the
3D case).

\end{document}